\documentclass[preprint, secnumarabic, amssymb, nobibnotes, aps, prb, superscriptaddress]{revtex4-1}

\usepackage{graphicx}
\usepackage{amsmath}

\begin{document}

\title{Nonlocal effects in plasmonic metasurfaces with almost touching surfaces}%

\author{Fan Yang}%
\email{f.yang16@imperial.ac.uk}
\affiliation{The Blackett Laboratory, Department of Physics, Imperial College London, London SW7 2AZ, UK}

\author{Emanuele Galiffi}
\affiliation{The Blackett Laboratory, Department of Physics, Imperial College London, London SW7 2AZ, UK}

\author{Paloma  Arroyo  Huidobro}
\affiliation{Instituto de Telecomunica{\c c}{\~o}es, Insituto Superior Tecnico-University of Lisbon, Avenida Rovisco Pais 1,1049-001 Lisboa, Portugal}

\author{John Pendry}
\affiliation{The Blackett Laboratory, Department of Physics, Imperial College London, London SW7 2AZ, UK}


\begin{abstract}
Geometrical singularities in plasmonic metasurfaces have recently been proposed for the enhancement of light-matter interactions, owing to their broadband light-harvesting properties and extreme plasmon confinement. However, the large plasmon momenta thus achieved lead to failure of local descriptions of the optical response of metals. Here we study a class of metasurfaces consisting of a periodic metal slab with a smooth modulation of its thickness. When the thinnest part shrinks, the two surfaces almost touch, forming a near-singular point. Using transformation optics, we show analytically how nonlocal effects, such as a blueshift of the resonance peaks and a reduced density of states, become important and cannot be ignored in this singular regime. The method developed in this paper is very general and can be used to model a variety of metasurfaces, providing valuable insight in the current context of ultra-thin plasmonic structures.
\end{abstract}

\maketitle

\section{Introduction}

The advancement of sophisticated nano-fabrication techniques has recently allowed for the realization of atomically thin films, structures with extremely sharp angles and touching points \cite{sondergaard2012plasmonic, ciraci2012probing, wiener2013electron, chikkaraddy2016single, benz2016single, chikkaraddy2016single,  maniyara2019tunable, boltasseva2019transdimensional}. The exotic behaviour of these singular structures hinges on small-scale geometrical features of nanometer and even sub-nanometer-scale, so that quantum effects, such as repulsion, diffusion and spill-out of electrons, become appreciable \cite{raza2011unusual, ciraci2012probing, scholl2012quantum,  mortensen2014generalized, toscano2015resonance}. In the case of noble metals, such as gold and silver, the nonlocal response is dominated by repulsion in the electron gas, resulting in size-dependent linewidth broadening and resonance shift. 

Singular metasurfaces constitute a special class of structured surfaces which feature sharp edges or ultra-small gaps. A conventional metasurface \cite{holloway2012overview, kildishev2013planar, yu2014flat, chen2016review} is characterized by two wave vectors, such that the selection of k-vectors is discretized. On the other hand, singular metasurfaces feature three wave vectors, owing to the additional length scale introduced by the singularity, which allows surface modes to exist over a continuum of quantum numbers \cite{pendry2017compacted,yang2018transformation,galiffi2019singular}. However, while a local singular structure has a broadband optical response, nonlocality sets a length scale that prevents the formation of a perfect singularity, thus yielding a discrete resonance spectrum. As recently pointed out in previous studies on singular surfaces, such as surfaces with a knife-edge profile \cite{yang2019nonlocal} and graphene-based gratings \cite{galiffi2019probing}, the degree of nonlocality determines the spacing of the resonance peaks in a far-field spectrum, so that singularities offer a valuable window into the microscopic physics of electrons in plasmonic materials.

In the past few years, the analytical technique of transformation optics (TO) \cite{ward1996refraction, leonhardt2006optical} has proven a powerful tool for the modelling of singular plasmonic structures both within the local approximation \cite{fernandez2012theory,aubry2010plasmonic} and including nonlocal effects as described by the hydrodynamic model \cite{fernandez2012transformation,yang2019nonlocal}. Here we deploy TO to model nonlocal effects of plasmonic metasurfaces in the form of thin gold gratings with almost touching surfaces, and explore their nonlocal response. Similar metasurfaces, in a non-singular regime, have been studied in the past \cite{kraft2015designing, pendry2019computing}. However, we have developed a different theoretical approach, which allows us to investigate the singular regime while accounting for nonlocal effects, which, as we show in the subsequency sections, become important for a near-singular metasurface, whose minimum thickness approaches the single-nanometre regime.

This paper is structured as follow: In section II, we present our analytical method, discussing in detail (1) the transformation of the geometry, source and dielectric function, (2) the dispersion relation and (3) the fields in real space and the metasurface reflection coefficient. In section III, we deploy our method in order to study a set of near-singular metasurfaces in terms of their far-field scattering spectra and field profiles, highlighting the contrast between local and nonlocal models. Finally, we close the Paper with conclusions in Secion IV.

\section{Principles}

\subsection{Transformation of geometry, source and dielectric function}

\begin{figure}[h]
\includegraphics[width=0.8\columnwidth]{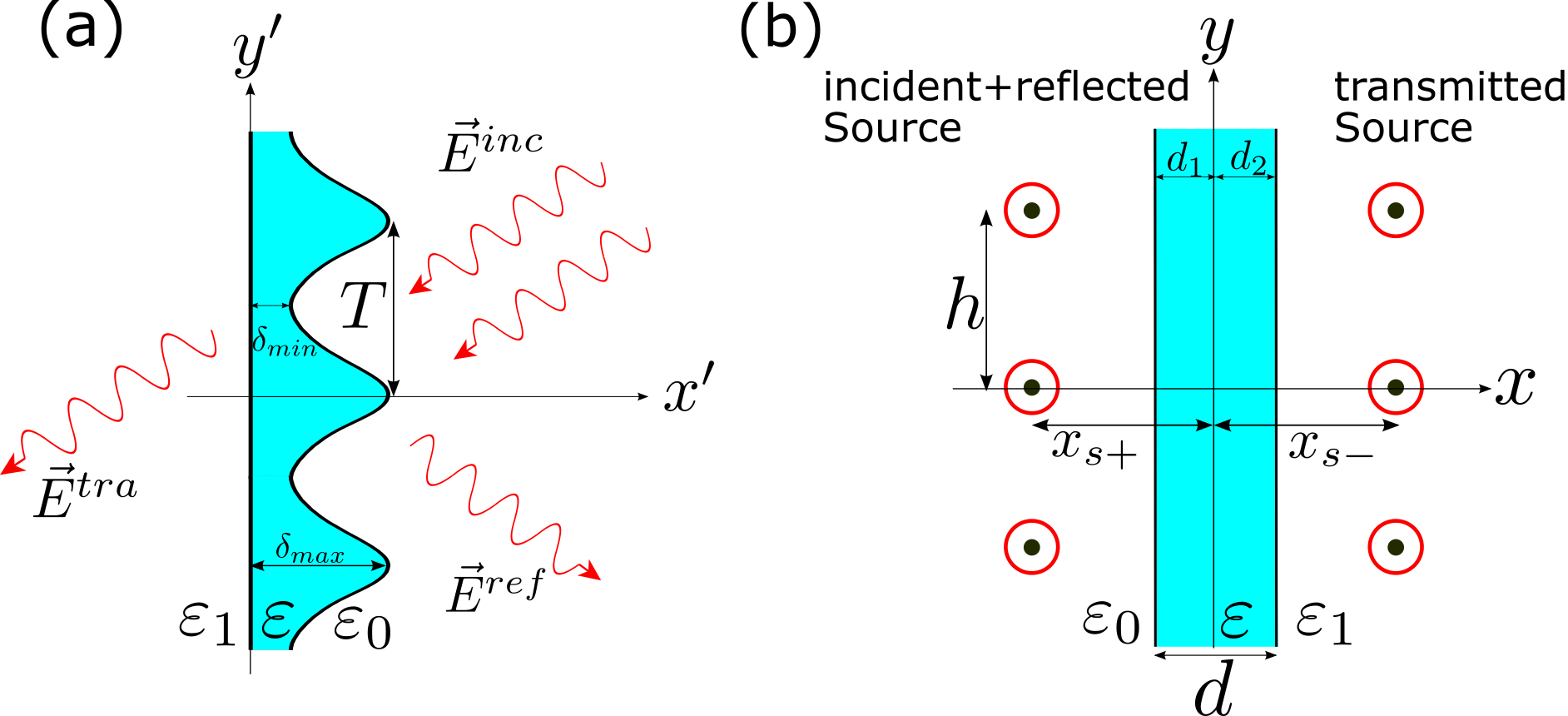}
\centering
\caption{Sketch showing the metasurface geometry and the source field in both frames. In the metasurface frame (a) the sources are plane waves (incident, reflected and transmitted waves), which we take to be generated by an array of magnetic line currents located at infinity. The sources are mapped into an array of magnetic line currents in the slab frame (b), where the structure is a flat slab.}
\label{SourceTransformation_sinusoidal}
\end{figure}

The metasurface studied in this paper is shown in Fig. \ref{SourceTransformation_sinusoidal}(a). It consists of a thin slab with periodic smooth thickness modulations and can be generated from the simple slab geometry in Fig. \ref{SourceTransformation_sinusoidal}(b) by using the transformation \cite{kraft2015designing}
\begin{equation}
\begin{split}
z^{'} &= \frac{T}{2\pi} \mathrm{ln} \bigg( \frac{1}{a e^{2\pi z /h}-w_0} - y_0 \bigg)\\
z &= \frac{h}{2\pi} \mathrm{ln} \bigg( \frac{1}{a (e^{2\pi z^{'} /T}+y_0)} + \frac{w_0}{a} \bigg)
\end{split}
\label{transformation}
\end{equation}
where $w_0 = \alpha a e^{-\frac{2\pi}{h}d_1}$, $y_0 = \alpha e^{-\frac{2\pi}{h}d}$, $d=d_1+d_2$ is the slab thickness and $h$ is the separation between line sources in the slab frame, which correspond to a linear array of monopoles. In this transformation, $z$ stands for the complex coordinate in the slab frame while $z^{'}$ for the coordinate in the metasurface frame. Also, $\delta_{min}$ and $\delta_{max}$ stand for the minimum and maximum grating thickness shown in Fig. \ref{SourceTransformation_sinusoidal}(a). Note that $T$, which acts as a scaling factor in the transformation affecting both the in-plane and out-of-plane components of the complex coordinate $z^{'}$, directly determines the period of the metasurface. In order to generate the metasurface in Fig. \ref{SourceTransformation_sinusoidal}(a), we also need $0<\alpha<1$. Moreover, without loss of generality, we place the two surfaces of the slab in the slab frame at $x=-d_1$ and $x = d_2$ with $d_1=d_2=d/2$, where $d$ is chosen as 1 throughout this paper. We additionally require that the flat interface of the metasurface lies on the $y$-axis, such that:
\begin{equation}
\begin{split}
a &= \frac{1}{e^{\frac{2\pi}{h}d_2}(1-\alpha^2 e^{-\frac{4\pi}{h}d})}
\end{split}
\end{equation}
Note that the wavy surface generated by Eq. \ref{transformation} is not sinusoidal, but nearly so when the modulation amplitude is not strong relative to the grating thickness.  

The transformation used also changes the form of the source excitation. In the metasurface frame, we consider an incident plane wave, as shown in Fig. \ref{SourceTransformation_sinusoidal}(a). For this near-sinusoidal metasurface, we follow the method introduced in \cite{yang2018transformation} to obtain the source representation, whereby incident, reflected and transmitted plane waves in the $z'$ frame correspond to monopole arrays in the $z$ frame. Note that incident and reflected waves on the right side of the metasurface correspond to the monopoles on the left side of the slab, and vice versa for transmitted waves, due to the coordinate inversion in Eq. \ref{transformation} [see Fig. \ref{SourceTransformation_sinusoidal}(b)].

With a lengthy but straightforward calculation (see Appendix A), the three types of waves in the slab frame are obtained and written as
\begin{equation}
\left\{
{\begin{array}{l}
H_z^{inc} = \int\limits_{-\infty}^{\infty} a_x \frac{e^{-|k_y||x-x_{s+}|}}{|k_y|} e^{i k_y y} d k_y +  \int\limits_{-\infty}^{\infty} a_y \text{sgn}(k_y) \frac{e^{-|k_y||x - x_{s+}|}}{\text{sgn}(x-x_{s+}) |k_y|} e^{i k_y y} d k_y\\
H_z^{ref} = -  \int\limits_{-\infty}^{\infty} r a_x \frac{e^{-|k_y||x-x_{s+}|}}{|k_y|} e^{i k_y y} d k_y +  \int\limits_{-\infty}^{\infty} r a_y \text{sgn}(k_y) \frac{e^{-|k_y||x - x_{s+}|}}{\text{sgn}(x-x_{s+}) |k_y|} e^{i k_y y} d k_y\\
H_z^{tra} = \int\limits_{-\infty}^{\infty} t \frac{k_{0x}^{'}}{k_{0x}} a_x \frac{e^{-|k_y||x-x_{s-}|}}{|k_y|} e^{i k_y y} d k_y - \int\limits_{-\infty}^{\infty} t a_y \text{sgn}(k_y) \frac{e^{-|k_y||x-x_{s-}|}}{\text{sgn}(x-x_{s-})|k_y|} e^{i k_y y} d k_y
\end{array}}
\right.
\label{source representation}
\end{equation}
where $x_{s+} = \frac{h}{2 \pi} \mathrm{ln} \frac{w_0}{a}$ is the $x$ coordinate of the monopole transformed from the incident and the reflected waves, while $x_{s-} = \frac{h}{2\pi} \mathrm{ln} \frac{1 + w_0 y_0}{a y_0}$ is that from the transmitted wave. In the above source representation, we have decomposed the source field into two parts: the mode denoted by $a_x$ corresponds to waves with the electric field parallel to the metasurface, while the electric field of the $a_y$ is normal to the metasurface. At normal incidence only the $a_x$ mode is excited \cite{yang2018transformation}.

\begin{figure}[h]
\includegraphics[width=0.7\columnwidth]{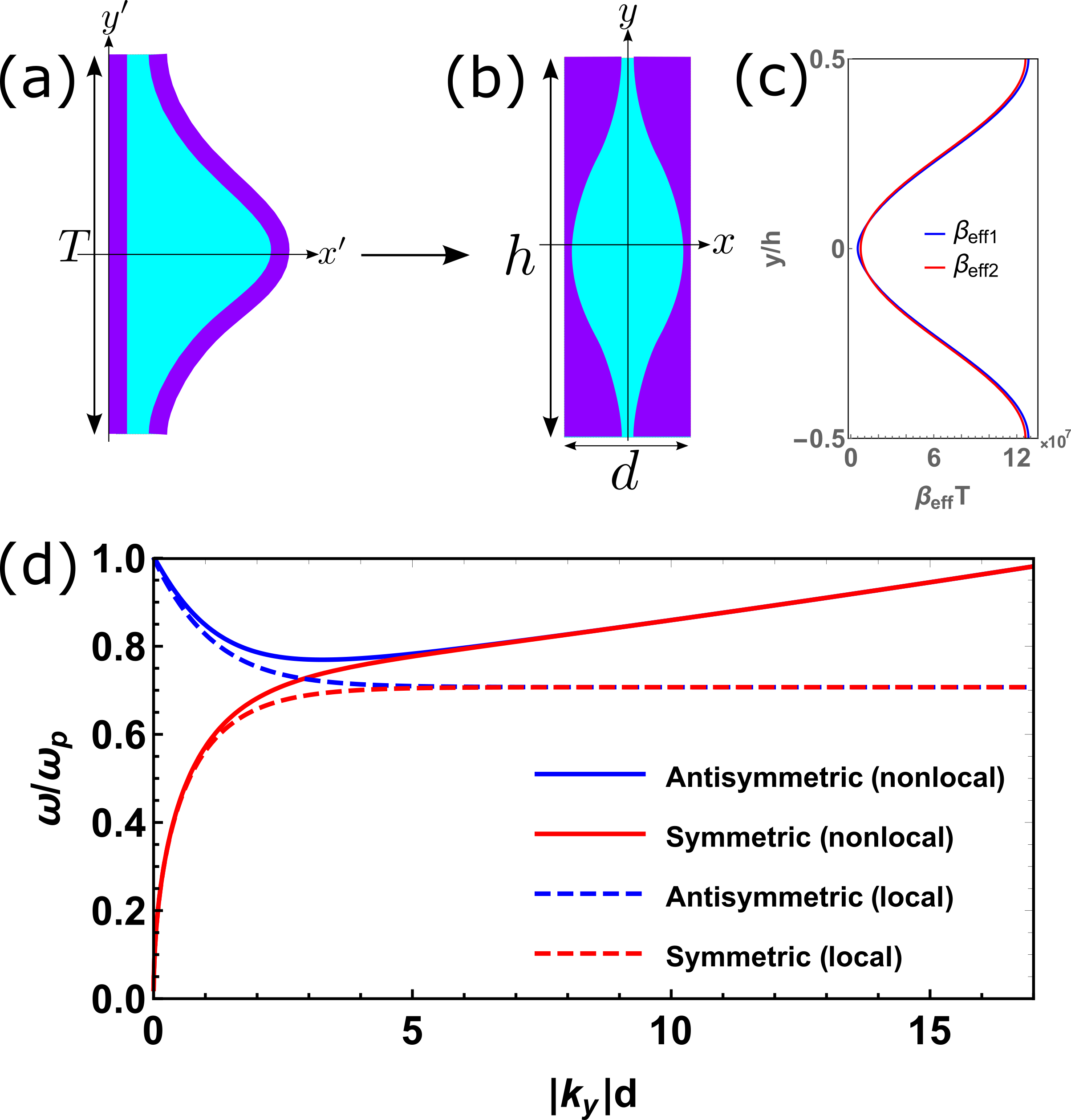}
\centering
\caption{Diagram for the longitudinal mode where the purple layer stands for the decay length of the longitudinal mode in (a) the metasurface frame and (b) the slab frame. The decay length is uniform in the metasurface frame, while after the transformation, the decay length varies along the slab. (c) The effective nonlocal parameter $\beta_{eff}$ as a function of $y$ at the two interfaces in the slab frame. (d) Dispersion relation for the near-sinusoidal metasurface, where the nonlocal dispersion relation ($\beta = 1.27\times 10^6$ m/s) is compared with the local case ($\beta=0$). The geometric parameters are $T=50 $ nm, the maximum gap $\delta_{max} = 10$ nm and the minimum gap $\delta_{min} = 0.5$ nm. The dispersion relation at $y=0.1h$ is illustrated.}
\label{Longitudinalmode_sinusoidal}
\end{figure}

The discrete nature of the electron gas determines the screening length ($\delta_C \sim0.1$ nm, for noble metals)\cite{kittel2004introduction} and prevents the electron density from diverging. This effect has been widely studied with the hydrodynamic model \cite{boardman1982electromagnetic, ciraci2012probing, raza2011unusual}. According to this model, both transverse and longitudinal modes are present inside the metal. The permittivity of the transverse mode, $\varepsilon_T$, is conserved under the conformal mapping between slab frame and metasurface frame. However, this symmetry does not hold for the longitudinal contribution to the dielectric function, whose form in the slab frame ($z = x + i y$) is related to that in the metasurface frame ($z^{'} = x^{'} + i y^{'}$) by \cite{fernandez2012transformation} 
\begin{equation}
\begin{split}
\varepsilon_L^{z}(\omega, \mathbf{k}, z) 
=  \varepsilon_L^{z^{'}}(\omega, \left|\frac{d z}{d z^{'}} \right|\mathbf{k})
=  \varepsilon_{\infty} - \frac{\omega_p^2}{\omega(\omega+i \Gamma)-\beta_{eff}^2(z)|\mathbf{k}|^2}
\end{split}
\end{equation}
where in the following calculation we choose $\varepsilon_{\infty}=1$, $\omega_p = 8.95$ eV/$\hbar$ and $\Gamma = 65.8$ meV/$\hbar$, which are typical parameters for gold \cite{novotny2012principles} and the effective nonlocal parameter, which acquires a spatial dependence in the slab frame, is
\begin{equation}
\begin{split}
\beta_{eff} (z) = \frac{h}{T} \left| \frac{y_0 (a e^{2\pi z / h} - w_0)^2 + w_0}{a e^{2\pi z / h}} -1 \right| \beta 
\end{split}
\end{equation} 
and $\beta = 1.27 \times 10^6$ m/s \cite{moreau2013impact}. 

The effective nonlocal parameter $\beta_{eff}$ on the two interfaces of the slab is depicted in Fig. \ref{Longitudinalmode_sinusoidal}(c). Note that the spatial deformation in the slab frame is not symmetric at the two interfaces ($x=-d_1$ and $x=d_2$), such that their effective nonlocal parameter $\beta_{eff}$ are not exactly the same. The blue line stands for the $\beta_{eff1}$ at the interface $x=-d_1$, while the red line is the $\beta_{eff2}$ at the interface $x=d_2$. However, the difference between these two $\beta_{eff}$ is much smaller than the variation of $\beta_{eff}$ along the slab. Therefore, in the following calculation, we use $(\beta_{eff1}+\beta_{eff2})/2$ as the effective $\beta$. 

In the presence of nonlocality, an additional longitudinal mode exists inside the metal. The decay length of this mode is uniform along the interface of metasurface, as shown in Fig. \ref{Longitudinalmode_sinusoidal}(a). However, upon transformation into the slab frame, the decay length of the longitudinal mode becomes periodically modulated along the slab (Fig. \ref{Longitudinalmode_sinusoidal}(b)). Note that the thinnest part of the metasurface is mapped to the part with largest decay length in the slab frame. In other words, nonlocal effects become more important near the minimum gap region.
 
\subsection{Dispersion relation}

In order to account for both the transverse and longitudinal modes we use two potential functions, the magnetic field $H_z$ and electric potential $\varphi$, in which $H_z$ corresponds to the transverse mode and $\varphi$ to the longitudinal mode  \cite{yang2019nonlocal}.

Using the source representation in Eq. \ref{source representation}, we can express the total field in $k$-space as
{\footnotesize
\begin{equation}
H_z(x,k_y)= \left\{{\begin{array}{lr}
\left( (1-r)a_x \frac{e^{-|k_y||x-x_{s+}|}}{|k_y|} + (1+r)a_y \text{sgn}(k_y)\frac{e^{-|k_y||x-x_{s+}|}}{\text{sgn}(x-x_{s+})|k_y|}+  b_+ e^{|k_y|x} \right) e^{i k_y y} ,&{x < -d/2}\\
\left( c_+ e^{|k_y|x} + c_- e^{-|k_y|x} \right) e^{i k_y y},&{-d/2 < x < d/2}\\
\left(  -t \frac{k_{0x}^{'}}{k_{0x}}  a_x \frac{e^{-|k_y||x-x_{s-}|}}{|k_y|} -t a_y \text{sgn}(k_y) \frac{e^{-|k_y||x-x_{s-}|}}{ \text{sgn}(x - x_{s-}) |k_y|} + b_- e^{-|k_y|x} \right) e^{i k_y y},&{x > d/2}
\end{array}} \right.
\label{field_kspace_Hz}
\end{equation}
}
and 
\begin{equation}
\begin{split}
\varphi (x,k_y) = \frac{\text{sgn}(k_y)}{\omega \varepsilon_0 \varepsilon} (d_+ e^{\kappa x} - d_- e^{-\kappa x}) e^{i k_y y}
\end{split}
\end{equation}
where $b_{\pm}$, $c_{\pm}$ and $d_{\pm}$ are the excited mode amplitudes, and $\kappa = \sqrt{k_y^2 + \frac{\omega_p^2}{\beta_{eff}} \frac{\varepsilon}{\varepsilon-1}}$ determines the decay of the longitudinal mode. The electric field components can be obtained as 
\begin{equation}
\begin{split}
E_x(k_x,y) &= \frac{i}{\omega \varepsilon}\frac{\partial H_z}{\partial y} - \frac{\partial \varphi}{\partial x} \\
E_y(k_x,y) &= -\frac{i}{\omega \varepsilon}\frac{\partial H_z}{\partial x} - \frac{\partial \varphi}{\partial y}
\end{split}
\end{equation}
in which $H_z$ contributes the divergence-free part of electric field, while $\varphi$ includes the curl-free part. Imposing the continuity of $H_z$ and $E_y$, together with vanishing normal component of current density $\mathbf{J}$ at the interface between the metal and the dielectric ($J_{x}=0$) \cite{ciraci2012probing,raza2011unusual}, we can obtain the excited mode amplitude. Furthermore, we take a WKB approximation \cite{griffiths2016introduction, fernandez2012transformation}, valid when the phase changes more rapidly than the amplitude (see Appendix C) \cite{yang2019nonlocal}. By looking at the pole of these mode amplitudes, we can obtain the dispersion relation given as Eq. \ref{DispersionRelation} in Appendix C. When the slab system is symmetric, i.e. $\varepsilon_1 = 1$ [$\varepsilon_1$ being the permittivity of the substrate of the metasurface, see Fig. 1(a)], the dispersion relation can be decomposed into anti-symmetric and symmetric modes. The dispersion relation for the anti-symmetric mode is 
\begin{equation}
\begin{split}
(\varepsilon-1)|k_y|(e^{|k_y|d}+1) \frac{e^{\kappa d}-1}{\kappa(e^{\kappa d}+1)} +  ((\varepsilon+1)e^{|k_y|d} + (\varepsilon-1)) = 0
\end{split}
\label{anti-symmetric}
\end{equation}
and for the symmetric mode
\begin{equation}
\begin{split}
(\varepsilon-1)|k_y|(e^{|k_y|d}-1) \frac{e^{\kappa d}+1}{\kappa(e^{\kappa d}-1)} + ((\varepsilon+1)e^{|k_y|d} - (\varepsilon-1)) = 0.
\end{split}
\label{symmetric}
\end{equation}
The dispersion relations are plotted in Fig. \ref{Longitudinalmode_sinusoidal}(d) for a metasurface parameterized by $T = 50$ nm, $\delta_{max}=10$ nm and $\delta_{min}=0.5$ nm. From Eqs. \ref{anti-symmetric} and \ref{symmetric}, it is clear that the dispersion relation reduces to the local case when $\kappa \rightarrow \infty$ ($\beta = 0$) \cite{aubry2010plasmonic}, which is depicted with dashed lines for comparison with the nonlocal dispersion relation (solid lines). 

From Fig. \ref{Longitudinalmode_sinusoidal}(d), we see that both local and nonlocal dispersion relations have two bands, the lower band representing the symmetric mode while the upper band the anti-symmetric one. However, the nonlocal dispersion relation differs from the local case in its asymptotic behavior. For the conventional local dispersion relation, the anti-symmetric and symmetric modes asymptotically approach $\omega_{sp}$. On the contrary, in a nonlocal model both bands asymptotically approach the longitudinal bulk mode $\omega = \sqrt{\beta_{eff}^2 k^2 + \omega_p^2}$ \cite{yang2019nonlocal}. In the large $k$ limit, $\omega \approx \beta_{eff} k$, giving rise to a linear dispersion relation, see the solid line in Fig. \ref{Longitudinalmode_sinusoidal}(d).

\subsection{Field in real space and reflection coefficient}

After obtaining the field in $k$-space, we can derive the field in real space by means of a Fourier transform. The potential functions, $H_z$ and $\varphi$, can then be expressed as follows. First, the magnetic field reads as,
{\footnotesize
\begin{equation}
\begin{split}
    H_z(x,y) 
    &= \left\{ {\begin{array}{lr}
    i 2\pi a \Gamma_{+} e^{\sqrt{k_{py}^2} x} (e^{i \int_{0}^{|y|} k_{py}(y') dy'} \pm e^{-i \int_{0}^{|y|} k_{py}(y') dy' + i \phi_0}) \frac{1}{1-e^{i \phi_0}}, & x<-d/2 \\
    i 2\pi a (\Lambda_{+} e^{\sqrt{k_{py}^2} x} + \Lambda_{-} e^{-\sqrt{k_{py}^2} x})(e^{i \int_{0}^{|y|} k_{py}(y') dy'} \pm e^{-i\int_{0}^{|y|} k_{py}(y') dy' + i \phi_0}) \frac{1}{1-e^{i \phi_0}}, & -d/2<x<d/2 \\ 
    i 2\pi a \Gamma_{-} e^{-\sqrt{k_{py}^2} x} (e^{i \int_{0}^{|y|} k_{py}(y') dy'} \pm e^{-i \int_{0}^{|y|} k_{py}(y') dy' + i \phi_0}) \frac{1}{1-e^{i \phi_0}}, & x>d/2
    \end{array}} \right.
\end{split}
\label{field real space Hz}
\end{equation}
}
where $\phi_0 = \int_{-h/2}^{h/2} k_{py}(y') dy'$. Second, the electric potential inside the metal is given by,
{\small
\begin{equation}
\begin{split}
   \varphi(x,y) = \frac{i 2\pi a \text{sgn}(k_{py})\text{sgn}(y)}{\omega \varepsilon_0 \varepsilon} (\Omega_+ e^{\kappa_p x} - \Omega_- e^{-\kappa_p x}) (e^{i \int_{0}^{|y|} k_{py}(y') dy'} \mp e^{-i\int_{0}^{|y|} k_{py}(y') dy' + i \phi_0}) \frac{1}{1-e^{i \phi_0}}
\end{split}
\label{field real space varphi}
\end{equation}
}
where coefficients $\Gamma_{\pm}$, $\Lambda_{\pm}$ and $\Omega_{\pm}$ are the mode amplitudes in real space.

From the potentials, the electric field can be calculated straightforwardly from Maxwell's equations. The energy loss of the system can then be obtained by calculating the power flow at the excitation point $y=0$. This power flow is then modeled as an effective surface conductivity as \cite{yang2018transformation}
\begin{equation}
\left\{ 
\begin{array}{lr}
   \sigma_{er}  = \sigma_{abs}^{x'} \sigma_{e0}  & a_x \; mode\\
   \sigma_{mr}  = \sigma_{abs}^{y'} \sigma_{m0} \sin^2{\theta_{in}} & a_y \; mode
\end{array}
\right.
\label{surface conductivity}
\end{equation}
in which $\sigma_{e0} = Z_0^{-1}$ and $\sigma_{m0} = Z_0$ are the free space electric and magnetic conductivity, $\theta_{in}$ is the incident angle and we give the expression of $\sigma_{abs}^{(x,y)'}$ in Appendix B and C. Note that Eq. \ref{surface conductivity} is just the real part of the surface conductivity, whose imaginary part can be derived using the Kramers-Kronig relations \cite{kittel2004introduction,jackson2012classical,dressel2005electrodynamics,yang2018transformation} 
\begin{equation}
\begin{split}
\sigma_{(e,m)i} = -\frac{1}{\pi}P\int\limits_{-\infty}^{\infty}\frac{\sigma_{(e,m)r}(s)}{s-\omega} ds = \frac{1}{\pi}  P \int\limits_{-\infty}^{\infty}\mathrm{ln}\big|s-\omega\big|\frac{d \sigma_{(e,m)r}(s)}{ds} ds
\end{split}
\label{Kramers-Kronig}
\end{equation} 

\begin{figure}[h]
\includegraphics[width=0.6\columnwidth]{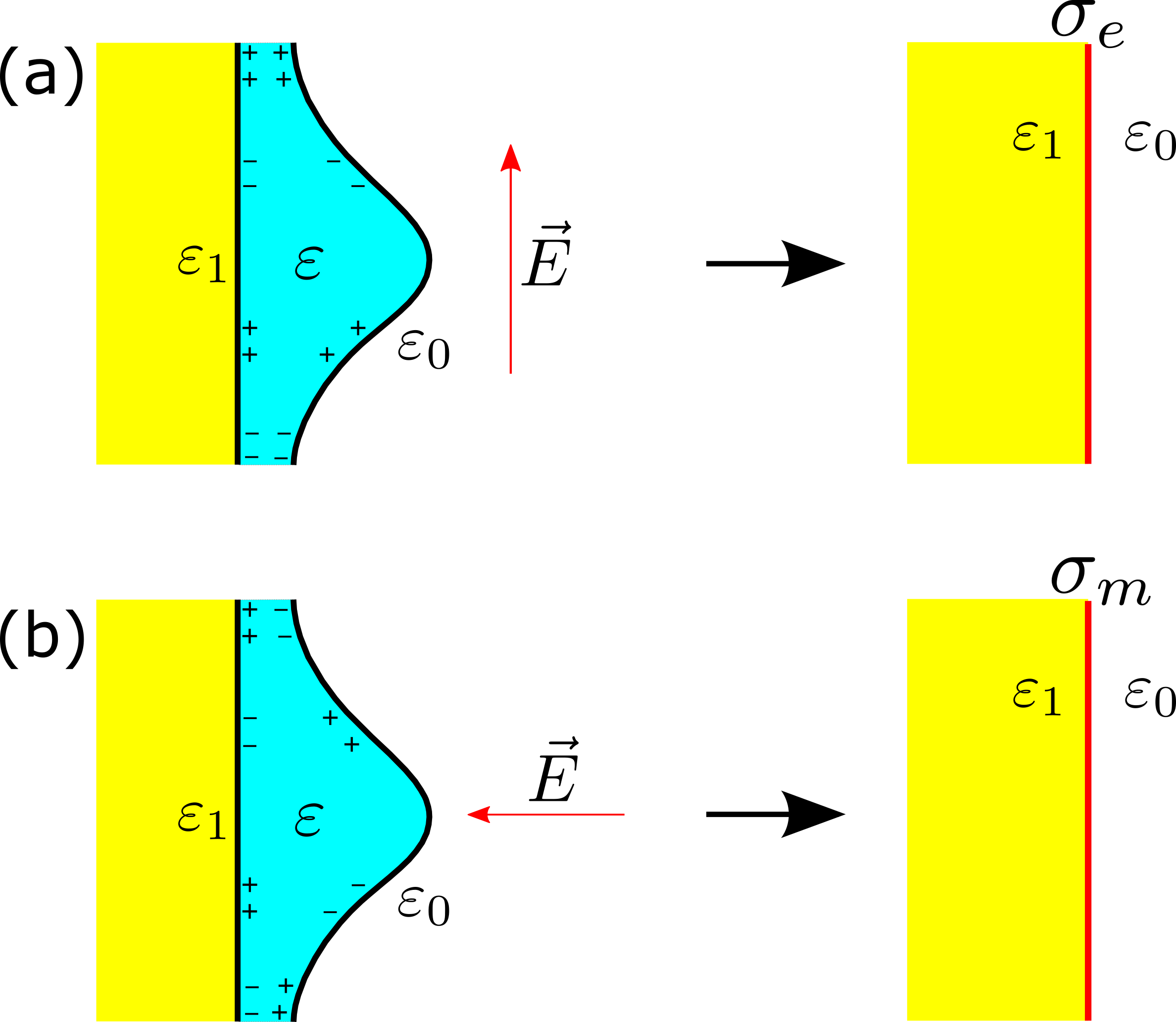}
\centering
\caption{Flat surface model. The energy dissipation by (a) $a_x$/ (b) $a_y$ mode is modeled as an effective electric/magnetic surface conductivity.}
\label{Flatsurfacemodel_sinusoidal}
\end{figure}

Using the flat surface model \cite{yang2018transformation}, the metasurface can be represented as a simple flat sheet with the electric and magnetic surface conductivities $(\sigma_e, \sigma_m)$, as shown in Fig. \ref{Flatsurfacemodel_sinusoidal}. In Fig. \ref{Flatsurfacemodel_sinusoidal}(a), the component of the electric field along the $y$ direction excites the $a_x$ mode, whose energy dissipation is modeled as an electric surface conductivity. In contrast, the component of electric field in the $x$ direction excites the $a_y$ mode, whose associated energy dissipation is equivalent to that of a magnetic surface conductivity.

With this simple flat surface geometry, the reflection and transmission coefficients can be readily obtained as
{\scriptsize
\begin{equation}
\begin{split}
        r=\frac{-4 \varepsilon_1  \sigma_m + 4 \sigma_e Z_0^2 \cos \theta_{in} \sqrt{\varepsilon_1 -\sin ^2\theta_{in}}-\sigma_e \sigma_m Z_0 \sqrt{\varepsilon_1 -\sin ^2\theta_{in}}+\varepsilon_1  \sigma_e \sigma_m Z_0 \cos \theta_{in}-4 Z_0 \sqrt{\varepsilon_1 -\sin ^2\theta_{in}}+4 \varepsilon_1  Z_0 \cos \theta_{in}}{4 \varepsilon_1  \sigma_m + 4 \sigma_e Z_0^2 \cos \theta_{in} \sqrt{\varepsilon_1 -\sin ^2\theta_{in}}+\sigma_e \sigma_m Z_0 \sqrt{\varepsilon_1 -\sin ^2\theta_{in}}+\varepsilon_1  \sigma_e \sigma_m Z_0 \cos\theta_{in}+4 Z_0 \sqrt{\varepsilon_1 -\sin ^2\theta_{in}}+4 \varepsilon_1  Z_0 \cos\theta_{in}}
\end{split}
\label{reflection_sinusoidal}
\end{equation}
} 
{\scriptsize
\begin{equation}
\begin{split}
        t=\frac{2 \varepsilon_1  Z_0 \cos \theta_{in} (4-\sigma_e \sigma_m)}{4 \varepsilon_1  \sigma_m + 4 \sigma_e Z_0^2 \cos \theta_{in} \sqrt{\varepsilon_1 -\sin ^2\theta_{in}}+\sigma_e \sigma_m Z_0 \sqrt{\varepsilon_1 -\sin ^2\theta_{in}}+\varepsilon_1  \sigma_e \sigma_m Z_0 \cos\theta_{in}+4 Z_0 \sqrt{\varepsilon_1 -\sin ^2\theta_{in}}+4 \varepsilon_1  Z_0 \cos\theta_{in}}
\end{split}
\label{transmission_sinusoidal}
\end{equation}
}
Under normal incidence ($\theta_{in}=0$), they can be written as
\begin{equation}
\begin{split}
    r &= \frac{\sqrt{\varepsilon_1} - 1 + \sigma_{e}/\sigma_{e0}}{\sqrt{\varepsilon_1} + 1 + \sigma_{e}/\sigma_{e0}} \\
    t &= \frac{2 \sqrt{\varepsilon_1} }{\sqrt{\varepsilon_1} + 1 + \sigma_{e}/\sigma_{e0}}
\end{split}
\end{equation} 

\section{Optical response of the metasurface}

\begin{figure}[h]
\includegraphics[width=0.9\columnwidth]{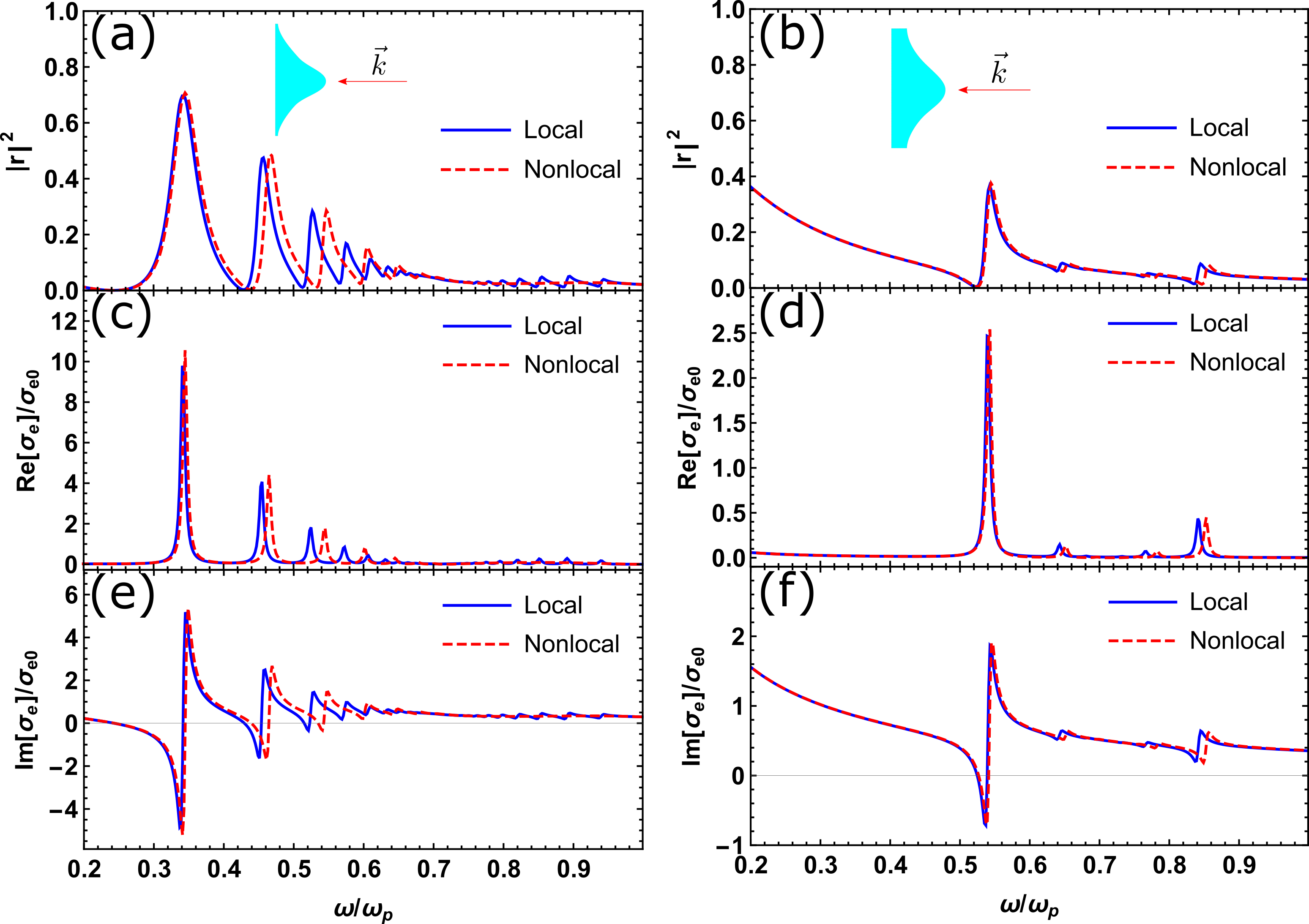}
\centering
\caption{The reflection spectrum $|r|^2$ (top row) and the real and imaginary components of the effective surface conductivity $\sigma_e$ (middle and bottom rows respectively) of the metasurface under normal incidence are compared in the near-singular case (left column) and in the non-singular one (right column). The geometric parameters are $T=50$ nm, the maximum gap $\delta_{max} = 10$ nm. The minimum gap $\delta_{min} = 0.5$ nm for the near-singular case, while $\delta_{min} = 5$ nm for the non-singular case.}
\label{ReflectionSpectrum}
\end{figure} 

In order to study how nonlocality influences the response of the system, the optical response of the metasurface is compared against the local approximation in Fig. \ref{ReflectionSpectrum}. The nonlocal parameter is taken as $\beta = 1.27 \times 10^6$ m/s as in Fig. \ref{Longitudinalmode_sinusoidal}, the grating period is $T = 50$ nm and the maximum thickness is $\delta_{max} = 10$ nm. Moreover, metasurfaces with different minimum gaps are considered in Fig. \ref{ReflectionSpectrum}, where the near-singular case on the left column ($\delta_{min} = 0.5$ nm) is compared with a non-singular case on the right column ($\delta_{min} = 5$ nm). Under normal incidence, we only have electric surface conductivity, shown in Figs. \ref{ReflectionSpectrum}(c)-(f). From the spectrum of surface conductivity, we can derive the reflection spectrum shown in Figs. \ref{ReflectionSpectrum}(a) and \ref{ReflectionSpectrum}(b). In order to check the accuracy of our analytic calculation, we carry out a comparison with exact numerical simulations implemented in the finite element solver Comsol Multiphysics. Excellent agreement between the two methods is shown in Fig. \ref{NumericalVerification} in Appendix D. 

From the comparison between local and nonlocal spectra of the near-singular case in the left column of Fig. \ref{ReflectionSpectrum}, we can observe a blue-shift of the peaks associated with the onset of nonlocality. Larger differences are observed in the high frequency regime, where the local calculation gives a discrete spectrum, while the nonlocal calculation has a continuous one. However, these differences reduce in the case of non-singular metasurface [Fig. \ref{ReflectionSpectrum} (b,d,f)], where the nonlocal result nearly coincides with the local solution. Therefore, nonlocal effects are significantly more pronounced when the metasurface is more singular.

\begin{figure}[h]
\includegraphics[width=1\columnwidth]{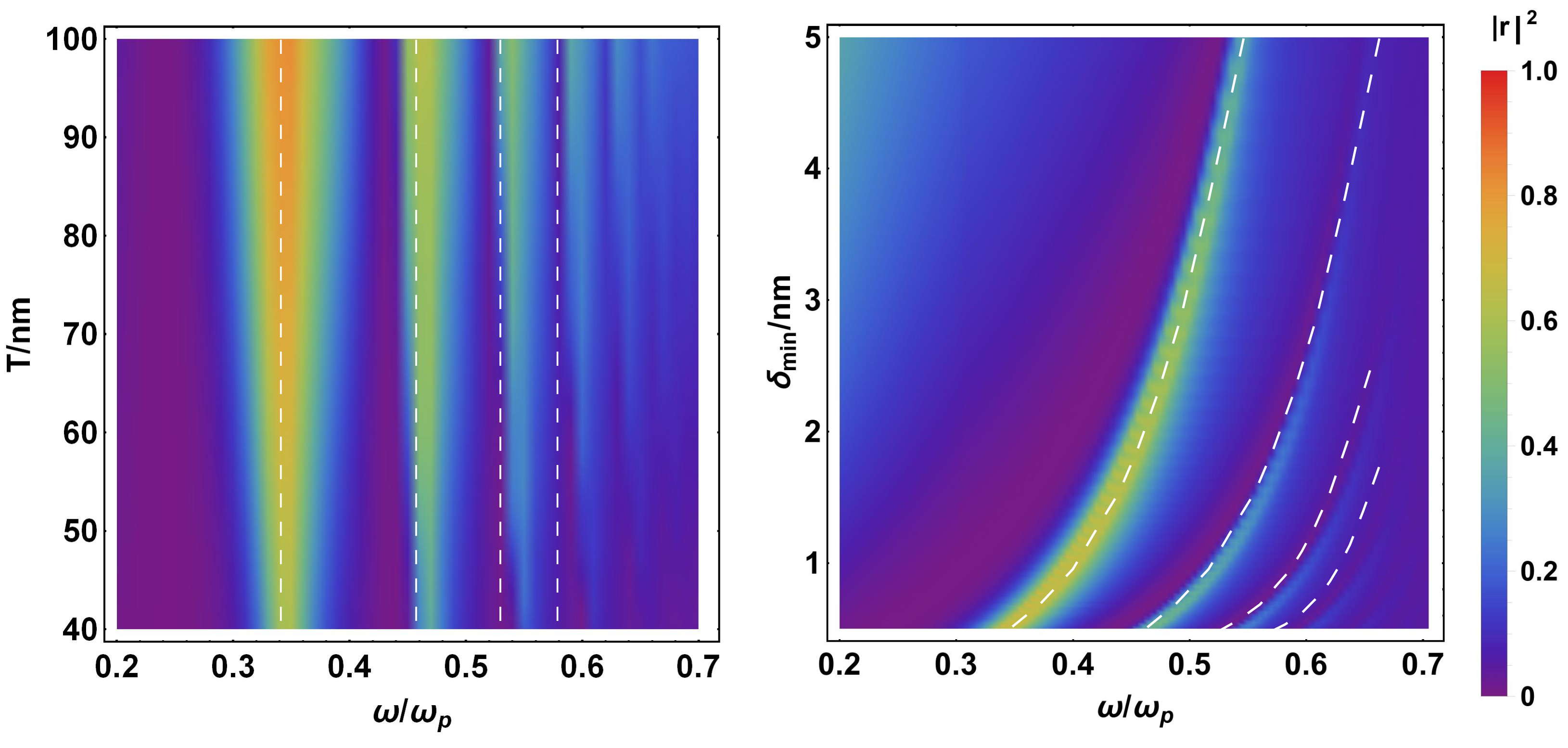}
\centering
\caption{Nonlocal effects in the far field spectrum. (a) Reflectance as a functioun of frequency and grating period $T$ for the metasurface under normal incidence. The geometric parameters are $h=23.6161 d$ and $\alpha = 0.8040$, which fixes the shape of metasurface when changing the grating period; (b) Reflectance as a function of frequency and minimum gap $\delta_{min}$ for the metasurface under normal incidence. The geometric parameters are $T=50$ nm and $\delta_{max}=10$ nm,  which is fixed when changing the minimum gap $\delta_{min}$. In both panels, the white dashed lines corresponds to the peak position of the local calculation.}
\label{ParameterPlot}
\end{figure} 

Nonlocality introduces a size-dependence in the response of the metasurface, otherwise scale invariant in the quasistatic limit. In Fig. \ref{ParameterPlot}(a), the dependence of reflection on the grating period $T$ and frequency are plotted for the nonlocal case. Since $T$ is a scaling factor, the shape of grating remains unchanged when $T$ varies. In the local model (white dashed lines), the subwavelength character of the grating implies that the response of the system is predominantly quasistatic, and therefore scale-invariant. However, the introduction of nonlocal effects induces the clear blueshift that appears for smaller grating sizes, and which is more pronounced for higher frequencies. We should point out that our analytic theory makes use of the quasi-static approximation, which typically results in a loss of accuracy for larger sizes due to coupling to radiation. However, radiative effects are not as relevant in singular plasmonic gratings as one may expect. While for localized plasmonic structures radiation damping becomes increasingly important for large particle sizes, which makes the quasistatic approximation inaccurate  \cite{fernandez2012transformation}, quasistatic calculations of singular plasmonic gratings with long periods yield very accurate results. The reason for this is that when the grating period increases, there are less periods per unit length, such that the total dipole moment does not increase \cite{yang2019nonlocal}. A quantitative comparison with numerical results for larger-period gratings is provided in Fig. \ref{NumericalVerification} in Appendix D, showing excellent agreement apart from minor discrepancies near the surface plasma frequency, for periods up to 100 nm.

We then investigate the geometric parameter $\delta_{min}$, which quantifies the proximity to the singular limit, and the resulting nonlocal effects. The contour plot in Fig. \ref{ParameterPlot}(b) shows reflection as a function of the gap size $\delta_{min}$ and frequency. When shrinking the gap size $\delta_{min}$, the number of visible peaks increases. This increase in resonances supported by the grating is a result of the hidden dimension at the singular point, which enables its spectral response to accommodate a higher density of states. In the local case, when $\delta_{min} \rightarrow 0$ these resonances will merge towards a continuum \cite{pendry2017compacted}. However, nonlocality sets a barrier to this continuous limit, saturating the density of states of the metasurface. By comparing the nonlocal spectrum with the local solution (white dashed lines), a saturation of the spectrum merging is observed for the smaller gap sizes $\delta_{min}$.

\begin{figure}[h]
\includegraphics[width=0.8\columnwidth]{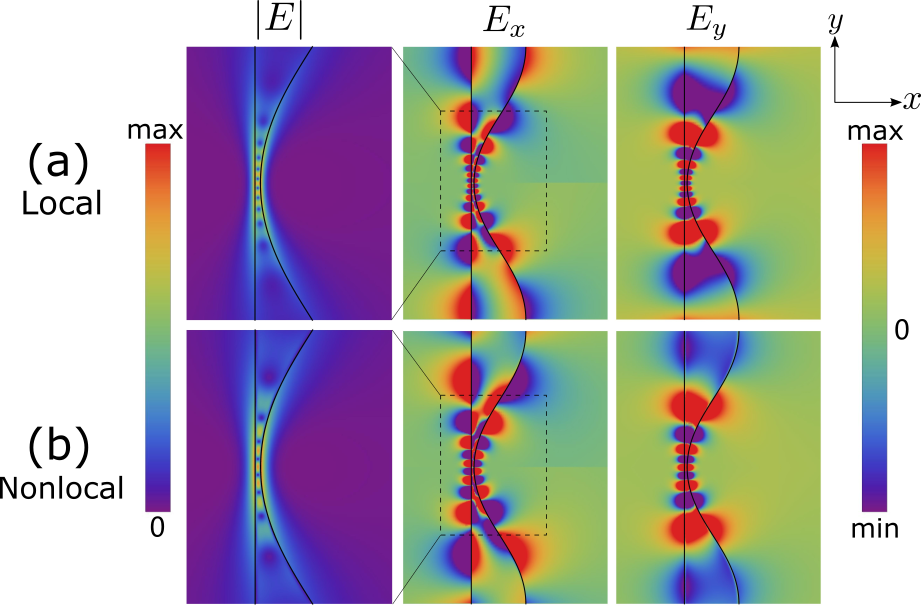}
\centering
\caption{The near field profile of a near-singular metasurface under normal incidence (a) without and (b) with nonlocality. The amplitude of the electric field $|E|$ (left column) and the $x$ and $y$ components $E_x$ and $E_y$ (center and right column respectively) are shown. The presence of nonlocality reduces the number of SPP oscillations in the field profile of $E_x$ and $E_y$. Note that we have zoomed in for the field $|E|$, whose maximum field enhancement factor is $\sim 45$ for the local case, which reduces to $\sim 40$ for the nonlocal one. The geometric parameters are $T=50$ nm, the maximum gap $\delta_{max} = 10$ nm and the minimum gap $\delta_{min} = 0.5$ nm.}
\label{Nearfield_sinusoidal}
\end{figure} 

We now turn our attention to the mode profiles, focusing on how the onset of nonlocality affects the near-field response. By substituting the reflection coefficient into the expressions for the fields in real space (Eqs. \ref{field real space Hz} and \ref{field real space varphi}), the field profile in the slab system is obtained. The resulting field distribution is subsequently obtained by mapping the field into the metasurface frame. Fig. \ref{Nearfield_sinusoidal} shows the  the near field response in the local case (panels (a)), and in the nonlocal one (panels (b)). The geometry of the metasurface studied here is parameterized by $T=50$ nm, the maximum gap $\delta_{max} = 10$ nm and the minimum gap $\delta_{min} = 0.5$ nm, which is the same as the near-singular metasurface in Fig. \ref{ReflectionSpectrum}. From the left to the right column, the colour plots show the amplitude of the electric field $|E|$, and its $x$ and $y$ components, $E_x$ and $E_y$ respectively. In order to compare the local and nonlocal plots, a working frequency $\omega = 0.65\omega_p$, below the surface plasmon frequency $\omega_{sp}$ is chosen, so that only the symmetric mode is excited. From this comparison, we see that the introduction of nonlocality leads to a reduced number of SPP oscillations, since nonlocality plays the role of smearing out the near-singular point, which effectively widens the minimum gap size. Furthermore, in the plot of $|E|$ we have zoomed in the region of interest, where the density of states is reduced as a result of nonlocality. This discrepancy will be more significant if we make the metasurface more singular by reducing the minimum gap size, effectively leading to saturation of the local density of states near the singular point.

\section{Conclusion}

In conclusion, we have studied how the response of a singular metasurface consisting of a smooth modulated metallic slab with nearly-touching surfaces is affected by nonlocality. We showed how transformation optics can be deployed to provide an accurate analytic description of repulsive quantum effects as described by the hydrodynamic model, which was shown to oppose the realisation of a geometrical singularity, thereby blueshifting the plasmonic bands of the metasurface. A convenient route towards this analytic solution was demonstrated, based on the representation of the grating as an effective current sheet, whose electric and magnetic surface conductivities were derived. This very general approach can in principle be applied to a wide range of metasurfaces, including surfaces with different kinds of singularities \cite{yang2019nonlocal}. 

From the comparison between the local and nonlocal calculation, we observed a blueshift of all resonance peaks and a reduced density of states. However, these nonlocal effects are only important for a near-singular metasurface, pointing out that these structures offer a convenient route for accessing nonlocal features of the electron gas via far-field measurements. Recent advancements in nanofabrication make ultra-small gaps ($\sim 1$ nm) possible to realize in practice \cite{maniyara2019tunable, boltasseva2019transdimensional}. In such small scale features, nonlocal effects cannot be ignored, and our theory is able to accurately account for these deviations.   

\section{Acknowledgements}
F.Y. acknowledges a Lee Family Scholarship for financial support. J.B.P. acknowledge funding from the Gordon and Betty Moore Foundation. E.G. acknowledges support through a studentship in the Centre for Doctoral Training on Theory and Simulation of Materials at Imperial College London funded by the EPSRC (EP/L015579/1). P.A.H. acknowledges funding from Funda\c{c}\~ao para a Ci\^encia e a Tecnologia and Instituto de Telecomunica\c c\~oes under projects CEECIND/03866/2017   and UID/EEA/50008/2019.

\appendix

\section{Source representation}

In this appendix, we present the derivation of the incident, reflected and transmitted waves in the slab frame. Starting from plane waves in the metasurface frame, we apply the transformation equation (Eq. \ref{transformation}) and write the source fields in the slab geometry. We assume that the metasurface is subwavelength and that the spatial region of interest satisfies $T \ll |x^{'}| \ll \lambda$. For the incident and reflected waves on the right-hand side of the metasurface ($T \ll x^{'} \ll \lambda$) under this assumption,  
\begin{equation}
\begin{split}
z  &= \frac{h}{2\pi} \left( \mathrm{ln} \left( \frac{1}{a (e^{2\pi z^{'} /T}+y_0)} + \frac{w_0}{a} \right)  \right) \\
& \approx \frac{h}{2\pi}  \bigg( \frac{e^{-\frac{2\pi}{T}z^{'}}}{w_0} + \mathrm{ln} \frac{w_0}{a} \bigg)
\end{split}
\end{equation}
Hence,
\begin{equation}
\begin{split}
z^{'} & \approx -\frac{T}{2\pi} \mathrm{ln} \bigg( \frac{2\pi w_0}{h}(z - \frac{h}{2 \pi} \mathrm{ln} \frac{w_0}{a}) \bigg)
\end{split}
\end{equation}
setting $x_{s+} = \frac{h}{2 \pi} \mathrm{ln} \frac{w_0}{a}$, we have 
\begin{equation}
\begin{split}
z^{'} & \approx -\frac{T}{2\pi} \mathrm{ln} \bigg( \frac{2\pi w_0}{h}(z - x_{s+}) \bigg)
\end{split}
\label{approx1}
\end{equation}

For the transmitted wave on the left side of singular surface we have $T \ll -x^{'} \ll \lambda$, so
\begin{equation}
\begin{split}
z &= \frac{h}{2\pi} \mathrm{ln} \bigg( \frac{1}{a y_0} (1 + \frac{1}{y_0}e^{\frac{2\pi}{T}z^{'}})^{-1} + \frac{w_0}{a} \bigg) \\
& \approx \frac{h}{2\pi} \mathrm{ln} \bigg( \frac{1}{a y_0} (1 - \frac{1}{y_0}e^{\frac{2\pi}{T}z^{'}}) + \frac{w_0}{a} \bigg) \\
& \approx \frac{h}{2\pi} \left( \mathrm{ln} \bigg( \frac{1 + w_0 y_0}{a y_0} \bigg) - \frac{e^{\frac{2\pi}{T}z^{'}}}{y_0(1 + w_0 y_0)}  \right)
\end{split}
\end{equation}
Hence,
\begin{equation}
\begin{split}
z^{'} & \approx \frac{T}{2\pi} \mathrm{ln} \bigg(  \frac{2\pi}{h}( \frac{h}{2\pi} \mathrm{ln} \frac{1 + w_0 y_0}{a y_0} - z ) y_0 (1 + w_0 y_0) \bigg)
\end{split}
\end{equation}
By setting $x_{s-} = \frac{h}{2\pi} \mathrm{ln} \frac{1 + w_0 y_0}{a y_0}$, we have 
\begin{equation}
\begin{split}
z^{'} & \approx \frac{T}{2\pi} \mathrm{ln} \bigg(  \frac{2\pi}{h}( x_{s-} - z ) y_0 (1 + w_0 y_0) \bigg)
\end{split}
\label{approx2}
\end{equation}

With Eqs. \ref{approx1} and \ref{approx2}, we have the source representation 
{\scriptsize
\begin{equation}
\begin{split}
H_z^{inc} &= H_0 e^{-i k_{0x} x^{'} + i k_{0y} y^{'}}\\
&\approx H_0 e^{- \frac{-i k_{0x} + k_{0y}}{2} \frac{T}{2\pi} \mathrm{ln} \big( \frac{\pi}{d}z \big)} \times e^{- \frac{-i k_{0x} - k_{0y}}{2} \frac{T}{2\pi} \mathrm{ln} \big( \frac{\pi}{d}z^* \big)}\\
&\approx H_0 \bigg(1 + \frac{i k_{0x} - k_{0y}}{2} \frac{T}{2\pi} \mathrm{ln} \big( \frac{2 \pi w_0}{h} (z -x_{s+} ) \big) \bigg) \times \bigg(1 + \frac{i k_{0x} + k_{0y}}{2} \frac{T}{2\pi} \mathrm{ln} \big( \frac{2 \pi w_0}{h} (z^* -x_{s+} ) \big) \bigg) \\
&= H_0 \bigg(1 + i\frac{k_{0x}T}{4\pi} \mathrm{ln} \big(\frac{2\pi w_0}{h} \big)^2 - i\frac{k_{0x}T}{4\pi} \int\limits_{-\infty}^{\infty} \frac{e^{-|k_y||x-x_{s+}|}}{|k_y|} e^{i k_y y} d k_y + \frac{k_{0y}T}{4\pi} \int\limits_{-\infty}^{\infty} \text{sgn}(k_y) \frac{e^{-|k_y||x-x_{s+}|}}{\text{sgn}(x-x_{s+})|k_y|} e^{i k_y y} d k_y \bigg)\\
&= H_0 \bigg(1 + i\frac{k_{0x}T}{4\pi} \mathrm{ln} \big(\frac{2\pi w_0}{h} \big)^2 \bigg) +  \int\limits_{-\infty}^{\infty} a_x \frac{e^{-|k_y||x-x_{s+}|}}{|k_y|} e^{i k_y y} d k_y +  \int\limits_{-\infty}^{\infty} a_y \text{sgn}(k_y) \frac{e^{-|k_y||x - x_{s+}|}}{\text{sgn}(x-x_{s+}) |k_y|} e^{i k_y y} d k_y
\end{split}   
\end{equation}
}
for the incident wave, where $a_x = - i\frac{k_{0x}T}{4\pi}H_0$ and $a_y = \frac{k_{0y}T}{4\pi}H_0$. For the reflected wave, we only need to change $k_{0x}$ into $-k_{0x}$ and obtain
{\scriptsize
\begin{equation}
\begin{split}
H_z^{ref} &= r H_0 e^{i k_{0x} x^{'} + i k_{0y} y^{'}}\\
&= r H_0 \bigg(1 - i\frac{k_{0x}T}{4\pi} \mathrm{ln} \big(\frac{2\pi w_0}{h} \big)^2 + i\frac{k_{0x}T}{4\pi} \int\limits_{-\infty}^{\infty} \frac{e^{-|k_y||x-x_{s+}|}}{|k_y|} e^{i k_y y} d k_y + \frac{k_{0y}T}{4\pi} \int\limits_{-\infty}^{\infty} \text{sgn}(k_y) \frac{e^{-|k_y||x-x_{s+}|}}{\text{sgn}(x-x_{s+})|k_y|} e^{i k_y y} d k_y \bigg)\\
&= r H_0 \bigg(1 - i\frac{k_{0x}T}{4\pi} \mathrm{ln} \big(\frac{2\pi w_0}{h} \big)^2 \bigg) -  \int\limits_{-\infty}^{\infty} r a_x \frac{e^{-|k_y||x-x_{s+}|}}{|k_y|} e^{i k_y y} d k_y +  \int\limits_{-\infty}^{\infty} r a_y \text{sgn}(k_y) \frac{e^{-|k_y||x - x_{s+}|}}{\text{sgn}(x-x_{s+}) |k_y|} e^{i k_y y} d k_y
\end{split}   
\end{equation}
}

For the transmitted wave, we have
{\scriptsize
\begin{equation}
\begin{split}
H_z^{tra} =& t H_0 e^{-i k_{0x}^{'} x^{'} + i k_{0y} y^{'}}\\
\approx & t H_0 e^{ \frac{-i k_{0x} + k_{0y}}{2} \frac{T}{2\pi} \mathrm{ln} \left(  \frac{2\pi}{h}( x_{s-} - z ) y_0 (1 + w_0 y_0) \right)} \times e^{ \frac{-i k_{0x} - k_{0y}}{2} \frac{T}{2\pi} \mathrm{ln} \left(  \frac{2\pi}{h}( x_{s-} - z^* ) y_0 (1 + w_0 y_0) \right)}\\
\approx & t H_0 \bigg(1 - \frac{i k_{0x}^{'} - k_{0y}}{2} \frac{T}{2\pi} \mathrm{ln} \left(  \frac{2\pi}{h}( x_{s-} - z ) y_0 (1 + w_0 y_0) \right) \bigg) \times \bigg(1 - \frac{i k_{0x}^{'} + k_{0y}}{2} \frac{T}{2\pi} \mathrm{ln} \left(  \frac{2\pi}{h}( x_{s-} - z^* ) y_0 (1 + w_0 y_0) \right) \bigg) \\
=& t H_0 \bigg(1 - i\frac{k_{0x}^{'} T}{4\pi} \mathrm{ln} \big( \frac{2\pi}{h} y_0 (1 + w_0 y_0) \big)^2 - i\frac{k_{0x}^{'} T}{4\pi} \mathrm{ln} |z - x_{s-}|^2  - \frac{k_{0y}T}{4\pi} \mathrm{ln} \big( \frac{z^*-x_{s-}}{z-x_{s-}} \big) \bigg) \\
=& t H_0 \bigg(1 - i\frac{k_{0x}^{'} T}{4\pi} \mathrm{ln} \big( \frac{2\pi}{h} y_0 (1 + w_0 y_0) \big)^2 \bigg) + \int\limits_{-\infty}^{\infty} t \frac{k_{0x}^{'}}{k_{0x}} a_x \frac{e^{-|k_y||x-x_{s-}|}}{|k_y|} e^{i k_y y} d k_y \\
& - \int\limits_{-\infty}^{\infty} t a_y \text{sgn}(k_y) \frac{e^{-|k_y||x-x_{s-}|}}{\text{sgn}(x-x_{s-})|k_y|} e^{i k_y y} d k_y
\end{split}   
\end{equation}
}

\section{Local calculation}

Without spatial dispersion in the metal, the field can be represented merely by a divergence-free magnetic field $H_z$, i.e. Eq. \ref{field_kspace_Hz}. By imposing the continuity condition at the two interfaces of $H_z$ and $E_y$, the field in k-space is obtained. In the calculation of the excited SPP mode in real space, we have two approaches: the discrete-k method and the continuous-k method. For the continuous-k method, we calculate the field generated by a single monopole and then sum the field of all monopoles together. In the discrete-k method, the periodicity of the monopole array only excites SPPs with a specific value of the k-vector ($k=n g$ where $n$ is an integer and $g = \frac{2\pi}{h}$ is the reciprocal lattice constant).

\subsubsection{Continuous-k method}
First, we consider two monopoles on the $x$-axis in the slab frame, one for incident and reflected waves at $z = x_{s+}$, and the other for transmitted wave at $z = x_{s-}$. Applying boundary conditions, we can obtain the mode amplitudes $b_{+,-}$ and $c_{+,-}$ in k-space, from which the field $H_z^0$ generated by the two monopoles in real space can be obtained by inverse Fourier transform.
\begin{equation}
\begin{split}
    H_z^0(x,y) &= \int\limits_{-\infty}^\infty H_z^0(x,k_y) e^{i{k_y}y} d{k_y}\\
    &= \left\{ {\begin{array}{lr}
    i 2\pi a \Gamma_{+} e^{\sqrt{k_{py}^2} x} e^{i k_{py} |y|}, & x<-d/2 \\
    i 2\pi a (\Lambda_{+} e^{\sqrt{k_{py}^2} x} + \Lambda_{-} e^{-\sqrt{k_{py}^2} x}) e^{i k_{py} |y|}, & -d/2<x<d/2 \\ 
    i 2\pi a \Gamma_{-} e^{-\sqrt{k_{py}^2} x} e^{i k_{py} |y|}, & x>d/2
    \end{array}} \right.
\end{split}
\end{equation}
where the coefficients $\Gamma_{+,-}$ and $\Lambda_{+,-}$ are the mode amplitudes in real space. Adding together the fields of all monopole pairs with ($x=x_{s+}$ and $x=x_{s-}$) at $y = n h$, we have
{\footnotesize
\begin{equation}
\begin{split}
    H_z(x,y) &=  \sum_n H_z^0(x,y-n h)\\
    &= \left\{ {\begin{array}{lr}
    i 2\pi a \Gamma_{+} e^{\sqrt{k_{py}^2} x} (e^{i k_{py} |y|} \pm e^{-i k_{py} |y| + i k_{py} h}) \frac{1}{1-e^{i k_{py} h}}, & x<-d/2 \\
    i 2\pi a (\Lambda_{+} e^{\sqrt{k_{py}^2} x} + \Lambda_{-} e^{-\sqrt{k_{py}^2} x})(e^{i k_{py} |y|} \pm e^{-i k_{py} |y| + i k_{py} h}) \frac{1}{1-e^{i k_{py} h}}, & -d/2<x<d/2 \\ 
    i 2\pi a \Gamma_{-} e^{-\sqrt{k_{py}^2} x} (e^{i k_{py} |y|} \pm e^{-i k_{py} |y| + i k_{py} h}) \frac{1}{1-e^{i k_{py} h}}, & x>d/2
    \end{array}} \right.
\end{split}
\end{equation}
}
where the positive sign is chosen for $a_x$, and the negative one for $a_y$. From Maxwell equations, the corresponding electric field can be calculated as
{\footnotesize
\begin{equation}
\begin{split}
    E_x(x,y) &= \frac{i}{\omega \varepsilon_0 \varepsilon} \frac{\partial H_z}{\partial y} \\
    &= \left\{ {\begin{array}{lr}
    -i \text{sgn}(y) \frac{2\pi a k_{py}}{\omega \varepsilon_0}  \Gamma_{+} e^{\sqrt{k_{py}^2} x} (e^{i k_{py} |y|} \mp e^{-i k_{py} |y| + i k_{py} h}) \frac{1}{1-e^{i k_{py} h}}, & x<-d/2 \\
   -i \text{sgn}(y) \frac{2\pi a k_{py}}{\omega \varepsilon_0 \varepsilon} (\Lambda_{+} e^{\sqrt{k_{py}^2} x} + \Lambda_{-} e^{-\sqrt{k_{py}^2} x}) (e^{i k_{py} |y|} \mp e^{-i k_{py} |y| + i k_{py} h}) \frac{1}{1-e^{i k_{py} h}}, & -d/2<x<d/2 \\ 
    -i \text{sgn}(y) \frac{2\pi a k_{py}}{\omega \varepsilon_0 \varepsilon_1} \Gamma_{-} e^{-\sqrt{k_{py}^2} x} (e^{i k_{py} |y|} \mp e^{-i k_{py} |y| + i k_{py} h}) \frac{1}{1-e^{i k_{py} h}}, & x>d/2
    \end{array}} \right.
\end{split}
\end{equation}
}
{\footnotesize
\begin{equation}
\begin{split}
    E_y(x,y) &= -\frac{i}{\omega \varepsilon_0 \varepsilon} \frac{\partial H_z}{\partial x} \\
    &= \left\{ {\begin{array}{lr}
     \frac{2\pi a \sqrt{k_{py}^2}}{\omega \varepsilon_0} \Gamma_{+} e^{\sqrt{k_{py}^2} x} (e^{i k_{py} |y|} \pm e^{-i k_{py} |y| + i k_{py} h}) \frac{1}{1-e^{i k_{py} h}}, & x<-d/2 \\
    \frac{2\pi a \sqrt{k_{py}^2}}{\omega \varepsilon_0 \varepsilon} (\Lambda_{+} e^{\sqrt{k_{py}^2} x} - \Lambda_{-} e^{-\sqrt{k_{py}^2} x})(e^{i k_{py} |y|} \pm e^{-i k_{py} |y| + i k_{py} h}) \frac{1}{1-e^{i k_{py} h}}, & -d/2<x<d/2 \\ 
    -\frac{2\pi a \sqrt{k_{py}^2}}{\omega \varepsilon_0 \varepsilon_1} \Gamma_{-} e^{-\sqrt{k_{py}^2} x} (e^{i k_{py} |y|} \pm e^{-i k_{py} |y| + i k_{py} h}) \frac{1}{1-e^{i k_{py} h}}, & x>d/2
    \end{array}} \right.
\end{split}
\end{equation}
}

To calculate the energy dissipated by excited SPP mode, we evaluate the power flow at $y=0_+$. For the $a_x$ mode, we have
\begin{equation}
\begin{split}
    H_z(x,y) = \left\{ {\begin{array}{lr}
    i 2\pi a_x \Gamma_{x+} e^{\sqrt{k_{py}^2} x} \frac{1+e^{i k_{py} h}}{1-e^{i k_{py} h}}, & x<-d/2 \\
    i 2\pi a_{x} (\Lambda_{x+} e^{\sqrt{k_{py}^2} x} + \Lambda_{x-} e^{-\sqrt{k_{py}^2} x}) \frac{1+e^{i k_{py} h}}{1-e^{i k_{py} h}}, & -d/2<x<d/2 \\ 
    i 2\pi a_{x} \Gamma_{x-} e^{-\sqrt{k_{py}^2} x} \frac{1+e^{i k_{py} h}}{1-e^{i k_{py} h}}, & x>d/2
    \end{array}} \right.
\end{split}
\end{equation}
The electric field $E_x$ is
\begin{equation}
\begin{split}
    E_x(x,y) = \left\{ {\begin{array}{lr}
    -i \frac{2\pi a_x k_{py}}{\omega \varepsilon_0}  \Gamma_{x+} e^{\sqrt{k_{py}^2} x}, & x<-d/2 \\
   -i \frac{2\pi a_x k_{py}}{\omega \varepsilon_0 \varepsilon} (\Lambda_{x+} e^{\sqrt{k_{py}^2} x} + \Lambda_{x-} e^{-\sqrt{k_{py}^2} x}), & -d/2<x<d/2 \\ 
    -i \frac{2\pi a_x k_{py}}{\omega \varepsilon_0 \varepsilon_1} \Gamma_{x-} e^{-\sqrt{k_{py}^2} x}, & x>d/2
    \end{array}} \right.
\end{split}
\end{equation}

Therefore, the total power absorbed by the $a_x$ mode is
\begin{equation}
\begin{split}
    P^{a_x}_{abs} &= 2 \int\limits_{-\infty}^{\infty} \frac{1}{2} \mathrm{Re}[S_y]\big|_{y=0_+} dx \\
    &= \frac{4 \pi^2 |a_x|^2}{\omega \varepsilon_0} \mathrm{Re} \bigg[ \left( \frac{1+e^{i k_{py} h}}{1-e^{i k_{py} h}} \right)^* 
    \bigg( k_{py} \frac{|\Gamma_{x+}|^2}{2 \mathrm{Re}[\sqrt{k_{py}^2}]} e^{-\mathrm{Re}[\sqrt{k_{py}^2}]d} + \frac{k_{py}}{\varepsilon_1} \frac{|\Gamma_{x-}|^2}{2 \mathrm{Re}[\sqrt{k_{py}^2}]} e^{-\mathrm{Re}[\sqrt{k_{py}^2}]d} \\
    &+ \frac{k_{py}}{\varepsilon} \big( \frac{|\Lambda_{x+}|^2}{2 \mathrm{Re}[\sqrt{k_{py}^2}]} (e^{\mathrm{Re}[\sqrt{k_{py}^2}]d} - e^{-\mathrm{Re}[\sqrt{k_{py}^2}]d}) + \frac{|\Lambda_{x-}|^2}{-2 \mathrm{Re}[\sqrt{k_{py}^2}]} (e^{-\mathrm{Re}[\sqrt{k_{py}^2}]d} - e^{\mathrm{Re}[\sqrt{k_{py}^2}]d}) \\
    &+ \frac{\Lambda_{x+} \Lambda_{x-}^{*}}{i 2 \mathrm{Im}[\sqrt{k_{py}^2}]} (e^{i \mathrm{Im}[\sqrt{k_{py}^2}]d} - e^{-i \mathrm{Im}[\sqrt{k_{py}^2}]d}) + \frac{\Lambda_{x-} \Lambda_{x+}^{*}}{-i 2 \mathrm{Im}[\sqrt{k_{py}^2}]} (e^{-i \mathrm{Im}[\sqrt{k_{py}^2}]d} - e^{i \mathrm{Im}[\sqrt{k_{py}^2}]d}) \big)
    \bigg)
    \bigg]
\end{split}
\end{equation}

For the $a_y$ mode at $y=0_+$, we have  
\begin{equation}
\begin{split}
    H_z(x,y) = \left\{ {\begin{array}{lr}
    i 2\pi a_y \Gamma_{y+} e^{\sqrt{k_{py}^2} x}, & x<-d/2 \\
    i 2\pi a_y (\Lambda_{y+} e^{\sqrt{k_{py}^2} x} + \Lambda_{y-} e^{-\sqrt{k_{py}^2} x}), & -d/2<x<d/2 \\ 
    i 2\pi a_y \Gamma_{y-} e^{-\sqrt{k_{py}^2} x}, & x>d/2
    \end{array}} \right.
\end{split}
\end{equation}
\begin{equation}
\begin{split}
    E_x(x,y) = \left\{ {\begin{array}{lr}
    -i \frac{2\pi a_y k_{py}}{\omega \varepsilon_0}  \Gamma_{y+} e^{\sqrt{k_{py}^2} x} \frac{1+e^{i k_{py} h}}{1-e^{i k_{py} h}}, & x<-d/2 \\
   -i \frac{2\pi a_y k_{py}}{\omega \varepsilon_0 \varepsilon} (\Lambda_{y+} e^{\sqrt{k_{py}^2} x} + \Lambda_{y-} e^{-\sqrt{k_{py}^2} x}) \frac{1+e^{i k_{py} h}}{1-e^{i k_{py} h}}, & -d/2<x<d/2 \\ 
    -i \frac{2\pi a_y k_{py}}{\omega \varepsilon_0 \varepsilon_1} \Gamma_{y-} e^{-\sqrt{k_{py}^2} x} \frac{1+e^{i k_{py} h}}{1-e^{i k_{py} h}}, & x>d/2
    \end{array}} \right.
\end{split}
\end{equation}
Therefore, the power absorbed by $a_y$ mode is
\begin{equation}
\begin{split}
    P^{a_y}_{abs}
    &= \frac{4 \pi^2 |a_y|^2}{\omega \varepsilon_0} \mathrm{Re} \bigg[ \left( \frac{1+e^{i k_{py} h}}{1-e^{i k_{py} h}} \right) 
    \bigg( k_{py} \frac{|\Gamma_{y+}|^2}{2 \mathrm{Re}[\sqrt{k_{py}^2}]} e^{-\mathrm{Re}[\sqrt{k_{py}^2}]d} + \frac{k_{py}}{\varepsilon_1} \frac{|\Gamma_{y-}|^2}{2 \mathrm{Re}[\sqrt{k_{py}^2}]} e^{-\mathrm{Re}[\sqrt{k_{py}^2}]d} \\
    &+ \frac{k_{py}}{\varepsilon} \big( \frac{|\Lambda_{y+}|^2}{2 \mathrm{Re}[\sqrt{k_{py}^2}]} (e^{\mathrm{Re}[\sqrt{k_{py}^2}]d} - e^{-\mathrm{Re}[\sqrt{k_{py}^2}]d}) + \frac{|\Lambda_{y-}|^2}{-2 \mathrm{Re}[\sqrt{k_{py}^2}]} (e^{-\mathrm{Re}[\sqrt{k_{py}^2}]d} - e^{\mathrm{Re}[\sqrt{k_{py}^2}]d}) \\
    &+ \frac{\Lambda_{y+} \Lambda_{y-}^{*}}{i 2 \mathrm{Im}[\sqrt{k_{py}^2}]} (e^{i \mathrm{Im}[\sqrt{k_{py}^2}]d} - e^{-i \mathrm{Im}[\sqrt{k_{py}^2}]d}) + \frac{\Lambda_{y-} \Lambda_{y+}^{*}}{-i 2 \mathrm{Im}[\sqrt{k_{py}^2}]} (e^{-i \mathrm{Im}[\sqrt{k_{py}^2}]d} - e^{i \mathrm{Im}[\sqrt{k_{py}^2}]d}) \big)
    \bigg)
    \bigg]
\end{split}
\end{equation}

By using the flat surface model \cite{yang2018transformation}, the power absorption of the $a_x$ and $a_y$ modes can be modeled as an effective surface conductivity, see Fig. \ref{Flatsurfacemodel_sinusoidal}. The component of the electric field along the $y$ direction excites the $a_x$ mode. For the near-sinusoidal metasurface in this paper, the $a_x$ mode does not have a well-defined symmetry and becomes a mixture of both anti-symmetric and symmetric modes. This is because the two boundaries of the metasurface are not symmetric in relation to the source, resulting in both the anti-symmetric and the symmetric modes being excited.  Likewise, the component of the electric field along the x direction excites $a_y$ mode, which is anti-symmetric dominantly. Following the same procedure we did for the wedge/groove singular metasurface, the effective surface conductivity is expressed as \cite{yang2018transformation}

\begin{equation}
\left\{ 
\begin{split}
\begin{array}{lr}
   \sigma_{er}  = \sigma_{abs}^{x'} \sigma_{e0}  &  a_x \; mode\\
   \sigma_{mr}  = \sigma_{abs}^{y'} \sigma_{m0} \sin^2{\theta_{in}} & a_y \; mode
\end{array}
\end{split}
\right.
\label{surface conductivity Re_sinusoidal}
\end{equation}
where $\sigma_{e0} = Z_0^{-1}$ and $\sigma_{m0} = Z_0$ are the free space electric and magnetic conductivity. The primed parameter in the above equation following the power-flow method reads
{\small
\begin{equation}
\begin{split}
   \sigma_{abs}^{(x,y)'}
    &= \frac{k_0 T}{2} \mathrm{Re} \bigg[ f(k_{py})
    \bigg( k_{py} \frac{|\Gamma_{(x,y)+}^{'}|^2}{2 \mathrm{Re}[\sqrt{k_{py}^2}]} e^{-\mathrm{Re}[\sqrt{k_{py}^2}]d} + \frac{k_{py}}{\varepsilon_1} \frac{|\Gamma_{y-}^{'}|^2}{2 \mathrm{Re}[\sqrt{k_{py}^2}]} e^{-\mathrm{Re}[\sqrt{k_{py}^2}]d} \\
    &+ \frac{k_{py}}{\varepsilon} \big( \frac{|\Lambda_{(x,y)+}^{'}|^2}{2 \mathrm{Re}[\sqrt{k_{py}^2}]} (e^{\mathrm{Re}[\sqrt{k_{py}^2}]d} - e^{-\mathrm{Re}[\sqrt{k_{py}^2}]d}) + \frac{|\Lambda_{(x,y)-}^{'}|^2}{-2 \mathrm{Re}[\sqrt{k_{py}^2}]} (e^{-\mathrm{Re}[\sqrt{k_{py}^2}]d} - e^{\mathrm{Re}[\sqrt{k_{py}^2}]d}) \\
    &+ \frac{\Lambda_{(x,y)+}^{'} \Lambda_{(x,y)-}^{'*}}{i 2 \mathrm{Im}[\sqrt{k_{py}^2}]} (e^{i \mathrm{Im}[\sqrt{k_{py}^2}]d} - e^{-i \mathrm{Im}[\sqrt{k_{py}^2}]d}) + \frac{\Lambda_{(x,y)-}^{'} \Lambda_{(x,y)+}^{'*}}{-i 2 \mathrm{Im}[\sqrt{k_{py}^2}]} (e^{-i \mathrm{Im}[\sqrt{k_{py}^2}]d} - e^{i \mathrm{Im}[\sqrt{k_{py}^2}]d}) \big)
    \bigg)
    \bigg]
\end{split}
\end{equation}
}
in which the primed coefficients $\Gamma_{(x,y)\pm}^{'}$ and $\Lambda_{(x,y)\pm}^{'}$ are related to $\Gamma_{(x,y)\pm}$ and $\Lambda_{(x,y)\pm}$ by \cite{yang2018transformation}
\begin{equation}
\begin{split}
\Gamma_{x\pm} &= (1-r)\Gamma_{x\pm}^{'} \\
\Lambda_{x\pm} &= (1-r)\Lambda_{x\pm}^{'} \\
\Gamma_{y\pm} &= (1+r)\Gamma_{y\pm}^{'} \\
\Lambda_{y\pm} &= (1+r)\Lambda_{y\pm}^{'} \\
\end{split}
\end{equation} 
and 
\begin{equation}
f(k_{py})=
\left\{ 
\begin{split}
\begin{array}{lr}
  \left( \frac{1+e^{i k_{py}h}}{1-e^{i k_{py}h}} \right)^*  &  a_x \; mode\\
  \frac{1+e^{i k_{py}h}}{1-e^{i k_{py}h}} & a_y \; mode
\end{array}
\end{split}
\right.
\end{equation}

\subsubsection{Discrete-k method}
If the field of one monopole is expressed as $H_z^0(x,y)$, the total field can be written as 
\begin{equation}
\begin{split}
H_z(x,y) &= \sum_n H_z^0(x,y-n h) \\
&= H_z^0(x,y) * \sum_n \delta(y - n h)
\end{split}
\end{equation}
where $n$ is an integer, $h$ is the period of the monopole source array and $*$ stands for convolution. Since $H_z(x,y)$ is a periodic function, it can be written as a Fourier series
\begin{equation}
\begin{split}
H_z(x,y) &= \sum_n c_n e^{i n g y}
\end{split}
\end{equation}
where
\begin{equation}
\begin{split}
c_n &= \frac{1}{h}\int_{-h/2}^{h/2} H_z(x,y) e^{-i n g y} dy \\
&= \frac{1}{h} \int_{-h/2}^{h/2} \sum_n H_z^0(x,y-n h) e^{-i n g y} dy\\
&= \frac{1}{h} \sum_n \int_{-h/2+ n h}^{h/2+ n h}  H_z^0(x,\tau_n) e^{-i n g (\tau_n + n h)} d\tau_n\\
&= \frac{1}{h} \sum_n \int_{-h/2+ n h}^{h/2+ n h}  H_z^0(x,\tau_n) e^{-i n g \tau_n} d\tau_n\\
&= \frac{1}{h} \int_{-\infty}^{\infty}  H_z^0(x,y) e^{-i n g y} dy\\
&= g H_z^0(x,n g)
\end{split}
\end{equation}
Therefore, 
\begin{equation}
\begin{split}
H_z(x,y) &= \sum_n g H_z^0(x,n g) e^{i n g y} \\
&= \int_{-\infty}^{\infty} g H_z^0(x,k_y) \sum_n \delta(k_y - n g) e^{i k_y y} d k_y
\end{split}
\end{equation}
Hence the field representation in $k$-space is $H_z^0(x,k_y) \times \sum_n g \delta(k_y - n g)$.

Solving the above equation system, we can calculate the field distribution of the excited mode in real space by inverse Fourier transform.
\begin{equation}
\begin{split}
    H_z(x,y) &= \int\limits_{-\infty}^\infty H_z(x,k_y) e^{i{k_y}y} d{k_y}\\
    &= \left\{ {\begin{array}{lr}
    a \sum_n g \Gamma_{n+} e^{|n|g x} e^{i n g y}, & x<-d/2 \\
    a \sum_n g (\Lambda_{n+} e^{|n| g x} + \Lambda_{n-} e^{-|n| g x}) e^{i n g y}, & -d/2<x<d/2 \\ 
    a \sum_n g \Gamma_{n-} e^{-|n|g x} e^{i n g y}, & x>d/2
    \end{array}} \right.
\end{split}
\end{equation}
where $\Gamma_{n\pm} = b_{\pm} \big|_{k_y= n g}$ and $\Lambda_{n\pm} = c_{\pm} \big|_{k_y= n g}$. Then the electric field can be calculated as
\begin{equation}
\begin{split}
    E_x(x,y) &= \frac{i}{\omega \varepsilon_0 \varepsilon} \frac{\partial H_z}{\partial y} \\
    &= \left\{ {\begin{array}{lr}
    -\frac{a}{\omega \varepsilon_0} \sum_n n g^2 \Gamma_{n+} e^{|n|g x} e^{i n g y}, & x<-d/2 \\
    -\frac{a}{\omega \varepsilon_0 \varepsilon} \sum_n n g^2 (\Lambda_{n+} e^{|n| g x} + \Lambda_{n-} e^{-|n| g x}) e^{i n g y}, & -d/2<x<d/2 \\ 
    -\frac{a}{\omega \varepsilon_0 \varepsilon_1} \sum_n n g^2 \Gamma_{n-} e^{-|n|g x} e^{i n g y}, & x>d/2
    \end{array}} \right.
\end{split}
\end{equation}

\begin{equation}
\begin{split}
    E_y(x,y) &= -\frac{i}{\omega \varepsilon_0 \varepsilon} \frac{\partial H_z}{\partial x} \\
    &= \left\{ {\begin{array}{lr}
    -\frac{i a}{\omega \varepsilon_0} \sum_n |n| g^2 \Gamma_{n+} e^{|n|g x} e^{i n g y}, & x<-d/2 \\
    -\frac{i a}{\omega \varepsilon_0 \varepsilon} \sum_n |n| g^2 (\Lambda_{n+} e^{|n| g x} - \Lambda_{n-} e^{-|n| g x}) e^{i n g y}, & -d/2<x<d/2 \\ 
    \frac{i a}{\omega \varepsilon_0 \varepsilon_1} \sum_n |n| g^2 \Gamma_{n-} e^{-|n|g x} e^{i n g y}, & x>d/2
    \end{array}} \right.
\end{split}
\end{equation}

The power absorption in one period is 
\begin{equation}
\begin{split}
    P_{abs} &= \int\limits_{slab} \frac{1}{2} \omega \varepsilon_0 \mathrm{Im}[\varepsilon] (|E_x|^2 + |E_y|^2) dx dy\\
    &= \frac{|a|^2 \mathrm{Im}[\varepsilon]}{\omega \varepsilon_0 |\varepsilon|^2} \sum_n n^2 g^4 h \int\limits_{-d/2}^{d/2} (|\Lambda_{n+}|^2 e^{2|n|g x} + |\Lambda_{n-}|^2 e^{-2|n|g x}) dx\\
    &= \frac{|a|^2 \mathrm{Im}[\varepsilon]}{\omega \varepsilon_0 |\varepsilon|^2} \sum_n n^2 g^4 h \left( \frac{|\Lambda_{n+}|^2}{2|n|g} (e^{|n| g d} - e^{-|n| g d}) + \frac{|\Lambda_{n-}|^2}{-2|n|g} (e^{-|n| g d} - e^{|n| g d}) \right)
\end{split}
\end{equation}
where the integration on $y$ cancels the terms $e^{i(n-n')g y}$ unless $n=n'$.
From the power absorption in one period, we can derive an effective surface conductivity as the continuous-$k$ method in Eq. \ref{surface conductivity Re_sinusoidal} where
\begin{equation}
\begin{split}
   \sigma_{abs}^{(x,y)'}
    &= \frac{k_0 T \mathrm{Im}[\varepsilon]}{8 \pi^2 |\varepsilon|^2} \sum_n n^2 g^4 h \left( \frac{|\Lambda_{n+}^{'}|^2}{2|n|g} (e^{|n| g d} - e^{-|n| g d}) + \frac{|\Lambda_{n-}^{'}|^2}{-2|n|g} (e^{-|n| g d} - e^{|n| g d}) \right)
\end{split}
\end{equation}

\begin{figure}[h]
\includegraphics[width=0.6\columnwidth]{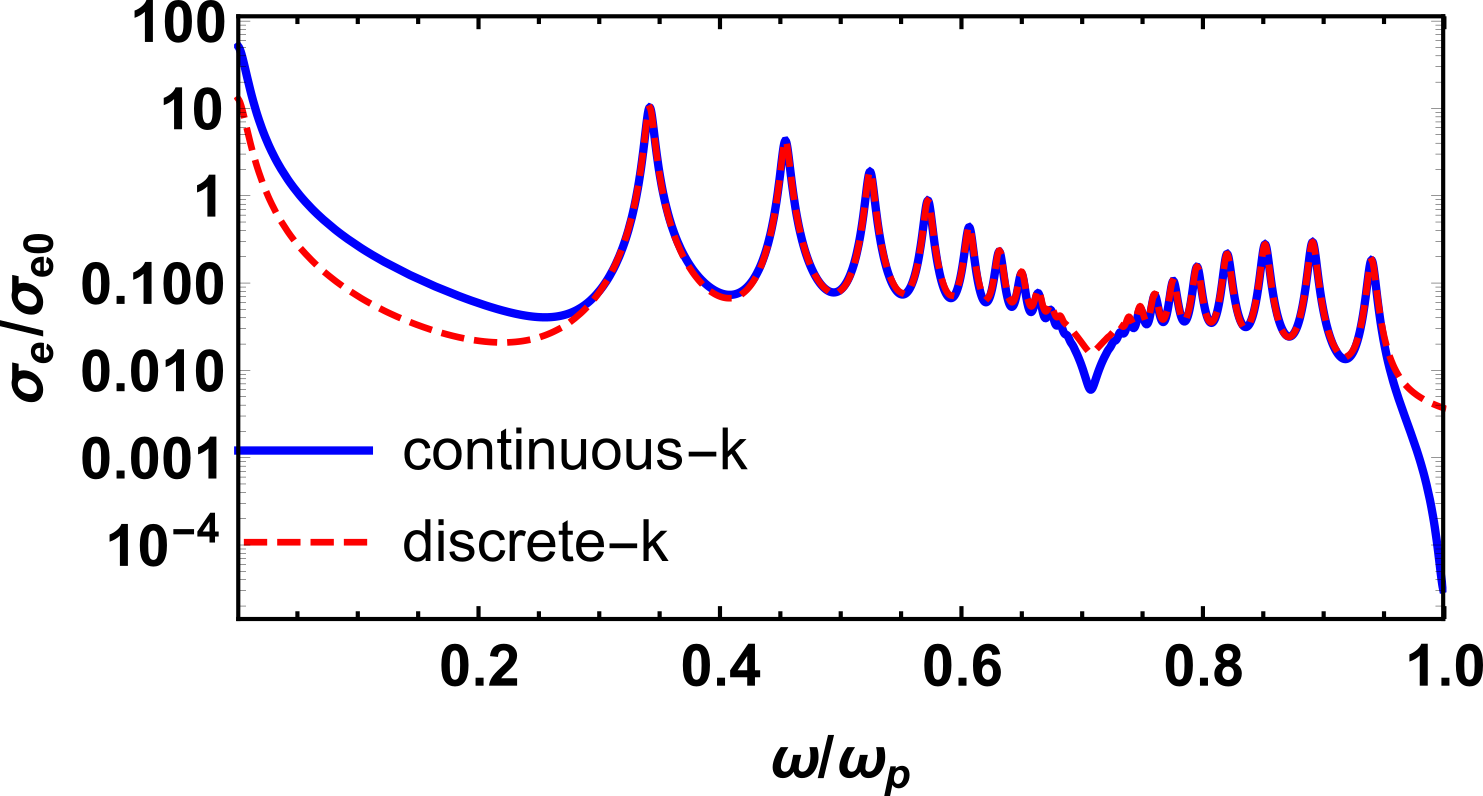}
\centering
\caption{Real part of the surface conductivity calculated with the discrete-$k$ and continuous-$k$ methods.}
\label{Surfaceconductivity_twomethod}
\end{figure}

The above two methods (discrete-$k$ and continuous-$k$) both give the spectrum of surface conductivity. They are identical when the k-vector of SPP is large (see Fig. \ref{Surfaceconductivity_twomethod}), but differ when the k-vector is small (near 0 and $\omega_p$ frequencies). This difference comes from neglecting the branch cut when applying the residue theorem in the continuous-k method. Therefore, the discrete-k method is more accurate for the local calculation.

\section{Nonlocal calculation}

In the presence of nonlocality, we have two potential functions: the divergence-free magnetic field $H_z$ and the curl-free scalar potential function $\varphi$. By imposing the continuity of $H_z$ and $E_y$, together with vanishing normal component of current density $\mathbf{J}$ at the metal interface ($J_x = 0$), we can obtain the mode amplitude in k-space. In the calculation, the WKB approximation is deployed. Under this approximation, the derivative of potential function on $y$ is only for the phase factor $e^{i k_y y}$. By looking at the pole of these mode amplitudes, we can arrive at the following dispersion relation:
{\footnotesize
\begin{equation}
\begin{split}
& \frac{|k_y|}{\kappa^2}(\varepsilon-1)((e^{2|k_y|d}-1)(e^{2\kappa d}-1)|k_y|(\varepsilon-1)\varepsilon_1 +  2 (1 + e^{2|k_y|d} + e^{2\kappa d} - 4 e^{(|k_y|+\kappa)d} + e^{2(|k_y|+\kappa)d})\varepsilon_1 \kappa  \\
& + (e^{2|k_y|d}-1)(e^{2\kappa d}+1)\varepsilon(\varepsilon_1+1)\kappa) +
4 e^{(|k_y|+\kappa)d} (\varepsilon(\varepsilon_1+1) \cosh[|k_y|d] + (\varepsilon^2+\varepsilon_1)\sinh[|k_y|d])\sinh[\kappa d] = 0
\end{split}
\label{DispersionRelation}
\end{equation}
}
Then by using a Fourier transform, the potential function $H_z$ and $\varphi$ in real space can be expressed as Eqs. \ref{field real space Hz} and \ref{field real space varphi}.

The continuous-k method allows for a more accurate way to study the variation of $k$ vector along the slab in the nonlocal calculation. However, neglecting the branch cut when using the residue theorem for the continuous-k method leads to some discrepancy when $k$ is small. In comparison, the discrete-k method has avoided the residue theorem such that there is no error in the case of the local calculation, but it is unable to describe the variation of $k$ vector along the slab in the nonlocal calculation. In view of the error of the continuous-k method and negligible nonlocal effect at low frequency, we utilize the local results from discrete-k method for low frequency and the nonlocal result from the continuous-k method for high frequency.

Following the scenario in the continuous-$k$ method, we evaluate the field at $y=0_+$ for the $a_x$ mode
\begin{equation}
\begin{split}
    H_z(x,y) = \left\{ {\begin{array}{lr}
    i 2\pi a_x \Gamma_{+}^{a_x} e^{\sqrt{k_{py}^2} x} \frac{1+e^{i \phi_0}}{1-e^{i \phi_0}}, & x<-d/2 \\
    i 2\pi a_x (\Lambda_{+}^{a_x} e^{\sqrt{k_{py}^2} x} + \Lambda_{-}^{a_x} e^{-\sqrt{k_{py}^2} x}) \frac{1+e^{\phi_0}}{1-e^{i \phi_0}}, & -d/2<x<d/2 \\ 
    i 2\pi a_x \Gamma_{-}^{a_x} e^{-\sqrt{k_{py}^2} x} \frac{1+e^{i \phi_0}}{1-e^{i \phi_0}}, & x>d/2
    \end{array}} \right.
\end{split}
\end{equation}
The electric field $E_x$ is
{\footnotesize
\begin{equation}
\begin{split}
    E_x(x,y) = \left\{ {\begin{array}{lr}
    -i \frac{2\pi a_x k_{py}}{\omega \varepsilon_0}  \Gamma_{+}^{a_x} e^{\sqrt{k_{py}^2} x}, & x<-d/2 \\
   -i \frac{2\pi a_x k_{py}}{\omega \varepsilon_0 \varepsilon} (\Lambda_{+}^{a_x} e^{\sqrt{k_{py}^2} x} + \Lambda_{-}^{a_x} e^{-\sqrt{k_{py}^2} x}) 
   -i \frac{2\pi a_x \text{sgn}(k_{py}) \kappa_p} {\omega \varepsilon_0 \varepsilon} (\Omega_{+}^{a_x} e^{\kappa_p x} + \Omega_{-}^{a_x} e^{-\kappa_p x}), & -d/2<x<d/2 \\
    -i \frac{2\pi a_x k_{py}}{\omega \varepsilon_0 \varepsilon_1} \Gamma_{-}^{a_x} e^{-\sqrt{k_{py}^2} x}, & x>d/2
    \end{array}} \right.
\end{split}
\end{equation}
}
For the $a_y$ mode at $y=0_+$, we have
\begin{equation}
\begin{split}
    H_z(x,y) = \left\{ {\begin{array}{lr}
    i 2\pi a_y \Gamma_{+}^{a_y} e^{\sqrt{k_{py}^2} x}, & x<-d/2 \\
    i 2\pi a_y (\Lambda_{+}^{a_y} e^{\sqrt{k_{py}^2} x} + \Lambda_{-}^{a_y} e^{-\sqrt{k_{py}^2} x}), & -d/2<x<d/2 \\ 
    i 2\pi a_y \Gamma_{-}^{a_y} e^{-\sqrt{k_{py}^2} x}, & x>d/2
    \end{array}} \right.
\end{split}
\end{equation}
{\scriptsize
\begin{equation}
\begin{split}
    E_x(x,y) = \left\{ {\begin{array}{lr}
    -i \frac{2\pi a_y k_{py}}{\omega \varepsilon_0}  \Gamma_{+}^{a_y} e^{\sqrt{k_{py}^2} x} \frac{1+e^{i \phi_0}}{1-e^{i \phi_0}}, & x<-d/2 \\
   (-i \frac{2\pi a_y k_{py}}{\omega \varepsilon_0 \varepsilon} (\Lambda_{+}^{a_y} e^{\sqrt{k_{py}^2} x} + \Lambda_{-}^{a_y} e^{-\sqrt{k_{py}^2} x}) 
   -i \frac{2\pi a_y \text{sgn}(k_{py}) \kappa_p}{\omega \varepsilon_0 \varepsilon} (\Omega_{+}^{a_y} e^{\kappa_p x} + \Omega_{-}^{a_y} e^{-\kappa_p x}) ) \frac{1+e^{i \phi_0}}{1-e^{i \phi_0}}, & -d/2<x<d/2 \\ 
    -i \frac{2\pi a_y k_{py}}{\omega \varepsilon_0 \epsilon_1} \Gamma_{-}^{a_y} e^{-\sqrt{k_{py}^2} x} \frac{1+e^{i \phi_0}}{1-e^{i \phi_0}}, & x>d/2
    \end{array}} \right.
\end{split}
\end{equation}
}

As opposed to the local case, the symmetric band and the anti-symmetric band overlap in frequencies between $\omega_{sp}$ and $\omega_p$. In this case, multiple SPP modes can be excited at the same frequency. We consider three poles together in the calculation of power flow, which can be written as

\begin{equation}
\begin{split}
P_{abs} &=  \frac{4 \pi^2 |a|^2}{\omega \varepsilon_0} \sum_{i,j=1,2,3} \mathrm{Re}\bigg[ f(k_{ij}) P_{abs}^{ij} 
\bigg]
\end{split}
\end{equation}
where $i$ and $j$ stand for the index of three modes considered. The expression for $P_{abs}^{ij}$ is 
{\footnotesize
\begin{equation}
\begin{split}
    & P^{a_x}_{i,j} \\
    &= \bigg( k_{pi} \frac{\Gamma_{+i} \Gamma_{+j}^{*}}{\sqrt{k_{pi}^2} + \sqrt{k_{pj}^2}^*} e^{-(\sqrt{k_{pi}^2} + \sqrt{k_{pj}^2}^*) \frac{d}{2}} + \frac{k_{pi}}{\varepsilon_1} \frac{\Gamma_{-i} \Gamma_{-j}^{*}}{\sqrt{k_{pi}^2} + \sqrt{k_{pj}^2}^*} e^{-(\sqrt{k_{pi}^2} + \sqrt{k_{pj}^2}^*) \frac{d}{2}} \\
    &+ \frac{k_{pi}}{\varepsilon} \big( \frac{\Lambda_{+i} \Lambda_{+j}^{*}}{\sqrt{k_{pi}^2} + \sqrt{k_{pj}^2}^*} (e^{(\sqrt{k_{pi}^2} + \sqrt{k_{pj}^2}^*)\frac{d}{2}} - e^{-(\sqrt{k_{pi}^2} + \sqrt{k_{pj}^2}^*)\frac{d}{2}}) + \frac{\Lambda_{-i} \Lambda_{+j}^{*}}{-\sqrt{k_{pi}^2}+\sqrt{k_{pj}^2}^*} (e^{(-\sqrt{k_{pi}^2}+\sqrt{k_{pj}^2}^*)\frac{d}{2}} - e^{-(-\sqrt{k_{pi}^2}+\sqrt{k_{pj}^2}^*)\frac{d}{2}}) \\
    &+ \frac{\Lambda_{+i} \Lambda_{-j}^{*}}{\sqrt{k_{pi}^2} - \sqrt{k_{pj}^2}^*} (e^{(\sqrt{k_{pi}^2} - \sqrt{k_{pj}^2}^*)\frac{d}{2}} - e^{-(\sqrt{k_{pi}^2} - \sqrt{k_{pj}^2}^*)\frac{d}{2}}) + \frac{\Lambda_{-i} \Lambda_{-j}^{*}}{-\sqrt{k_{pi}^2}-\sqrt{k_{pj}^2}^*} (e^{(-\sqrt{k_{pi}^2}-\sqrt{k_{pj}^2}^*)\frac{d}{2}} - e^{-(-\sqrt{k_{pi}^2}-\sqrt{k_{pj}^2}^*)\frac{d}{2}})
    \big) \\
    &+ \frac{\text{sgn}(k_{pi})\kappa_{pi}}{\varepsilon} \big( \frac{\Omega_{+i} \Lambda_{+j}^{*}}{\kappa_{pi} + \sqrt{k_{pj}^2}^*} (e^{(\kappa_{pi} + \sqrt{k_{pj}^2}^*)\frac{d}{2}} - e^{-(\kappa_{pi} + \sqrt{k_{pj}^2}^*)\frac{d}{2}}) + \frac{\Omega_{-i} \Lambda_{+j}^{*}}{-\kappa_{pi}+\sqrt{k_{pj}^2}^*} (e^{(-\kappa_{pi}+\sqrt{k_{pj}^2}^*)\frac{d}{2}} - e^{-(-\kappa_{pi}+\sqrt{k_{pj}^2}^*)\frac{d}{2}}) \\
    &+ \frac{\Omega_{+i} \Lambda_{-j}^{*}}{\kappa_{pi} - \sqrt{k_{pj}^2}^*} (e^{(\kappa_{pi} - \sqrt{k_{pj}^2}^*)\frac{d}{2}} - e^{-(\kappa_{pi} - \sqrt{k_{pj}^2}^*)\frac{d}{2}}) + \frac{\Omega_{-i} \Lambda_{-j}^{*}}{-\kappa_{pi}-\sqrt{k_{pj}^2}^*} (e^{(-\kappa_{pi}-\sqrt{k_{pj}^2}^*)\frac{d}{2}} - e^{-(-\kappa_{pi}-\sqrt{k_{pj}^2}^*)\frac{d}{2}})
    \big)
    \bigg)
\end{split}
\end{equation}
}
and
\begin{equation}
f(k_{ij})=
\left\{ 
\begin{split}
\begin{array}{lr}
  \left( \frac{1+e^{i \phi_{0j}}}{1-e^{i \phi_{0j}}} \right)^*  &  a_x \; mode\\
  \frac{1+e^{i \phi_{0i}}}{1-e^{i \phi_{0i}}} & a_y \; mode
\end{array}
\end{split}
\right.
\end{equation} 
Then, this power flow can be modeled as an effective surface conductivity, written as
\begin{equation}
\left\{ 
\begin{split}
\begin{array}{lr}
   \sigma_{er}  = \sigma_{abs}^{x'} \sigma_{e0}  & a_x \; mode\\
   \sigma_{mr}  = \sigma_{abs}^{y'} \sigma_{m0} \sin^2{\theta_{in}} & a_y \; mode
\end{array}
\end{split}
\right.
\end{equation}
in which 
\begin{equation}
\begin{split}
\sigma_{abs}^{(x,y)'} 
&= \frac{k_0 T}{2} \sum_{i,j=1,2,3} \mathrm{Re} \bigg[f(k_{ij}) \sigma_{abs}^{ij}
\bigg]
\end{split}
\end{equation}
where $\sigma_{abs}^{ij}$ is written as
{\footnotesize
\begin{equation}
\begin{split}
    & \sigma_{abs}^{ij}  \\
    &= \bigg( k_{pi} \frac{\Gamma_{+i}^{'} \Gamma_{+j}^{'*}}{\sqrt{k_{pi}^2} + \sqrt{k_{pj}^2}^*} e^{-(\sqrt{k_{pi}^2} + \sqrt{k_{pj}^2}^*) \frac{d}{2}} + \frac{k_{pi}}{\varepsilon_1} \frac{\Gamma_{-i}^{'} \Gamma_{-j}^{'*}}{\sqrt{k_{pi}^2} + \sqrt{k_{pj}^2}^*} e^{-(\sqrt{k_{pi}^2} + \sqrt{k_{pj}^2}^*) \frac{d}{2}} \\
    &+ \frac{k_{pi}}{\varepsilon} \big( \frac{\Lambda_{+i}^{'} \Lambda_{+j}^{'*}}{\sqrt{k_{pi}^2} + \sqrt{k_{pj}^2}^*} (e^{(\sqrt{k_{pi}^2} + \sqrt{k_{pj}^2}^*)\frac{d}{2}} - e^{-(\sqrt{k_{pi}^2} + \sqrt{k_{pj}^2}^*)\frac{d}{2}}) + \frac{\Lambda_{-i}^{'} \Lambda_{+j}^{'*}}{-\sqrt{k_{pi}^2}+\sqrt{k_{pj}^2}^*} (e^{(-\sqrt{k_{pi}^2}+\sqrt{k_{pj}^2}^*)\frac{d}{2}} - e^{-(-\sqrt{k_{pi}^2}+\sqrt{k_{pj}^2}^*)\frac{d}{2}}) \\
    &+ \frac{\Lambda_{+i}^{'} \Lambda_{-j}^{'*}}{\sqrt{k_{pi}^2} - \sqrt{k_{pj}^2}^*} (e^{(\sqrt{k_{pi}^2} - \sqrt{k_{pj}^2}^*)\frac{d}{2}} - e^{-(\sqrt{k_{pi}^2} - \sqrt{k_{pj}^2}^*)\frac{d}{2}}) + \frac{\Lambda_{-i}^{'} \Lambda_{-j}^{'*}}{-\sqrt{k_{pi}^2}-\sqrt{k_{pj}^2}^*} (e^{(-\sqrt{k_{pi}^2}-\sqrt{k_{pj}^2}^*)\frac{d}{2}} - e^{-(-\sqrt{k_{pi}^2}-\sqrt{k_{pj}^2}^*)\frac{d}{2}})
    \big) \\
    &+ \frac{\text{sgn}(k_{pi})\kappa_{pi}}{\varepsilon} \big( \frac{\Omega_{+i}^{'} \Lambda_{+j}^{'*}}{\kappa_{pi} + \sqrt{k_{pj}^2}^*} (e^{(\kappa_{pi} + \sqrt{k_{pj}^2}^*)\frac{d}{2}} - e^{-(\kappa_{pi} + \sqrt{k_{pj}^2}^*)\frac{d}{2}}) + \frac{\Omega_{-i}^{'} \Lambda_{+j}^{'*}}{-\kappa_{pi}+\sqrt{k_{pj}^2}^*} (e^{(-\kappa_{pi}+\sqrt{k_{pj}^2}^*)\frac{d}{2}} - e^{-(-\kappa_{pi}+\sqrt{k_{pj}^2}^*)\frac{d}{2}}) \\
    &+ \frac{\Omega_{+i}^{'} \Lambda_{-j}^{'*}}{\kappa_{pi} - \sqrt{k_{pj}^2}^*} (e^{(\kappa_{pi} - \sqrt{k_{pj}^2}^*)\frac{d}{2}} - e^{-(\kappa_{pi} - \sqrt{k_{pj}^2}^*)\frac{d}{2}}) + \frac{\Omega_{-i}^{'} \Lambda_{-j}^{'*}}{-\kappa_{pi}-\sqrt{k_{pj}^2}^*} (e^{(-\kappa_{pi}-\sqrt{k_{pj}^2}^*)\frac{d}{2}} - e^{-(-\kappa_{pi}-\sqrt{k_{pj}^2}^*)\frac{d}{2}})
    \big)
    \bigg)
\end{split}
\end{equation}
}

\section{Numerical verification}

\begin{figure}[h]
\includegraphics[width=0.9\columnwidth]{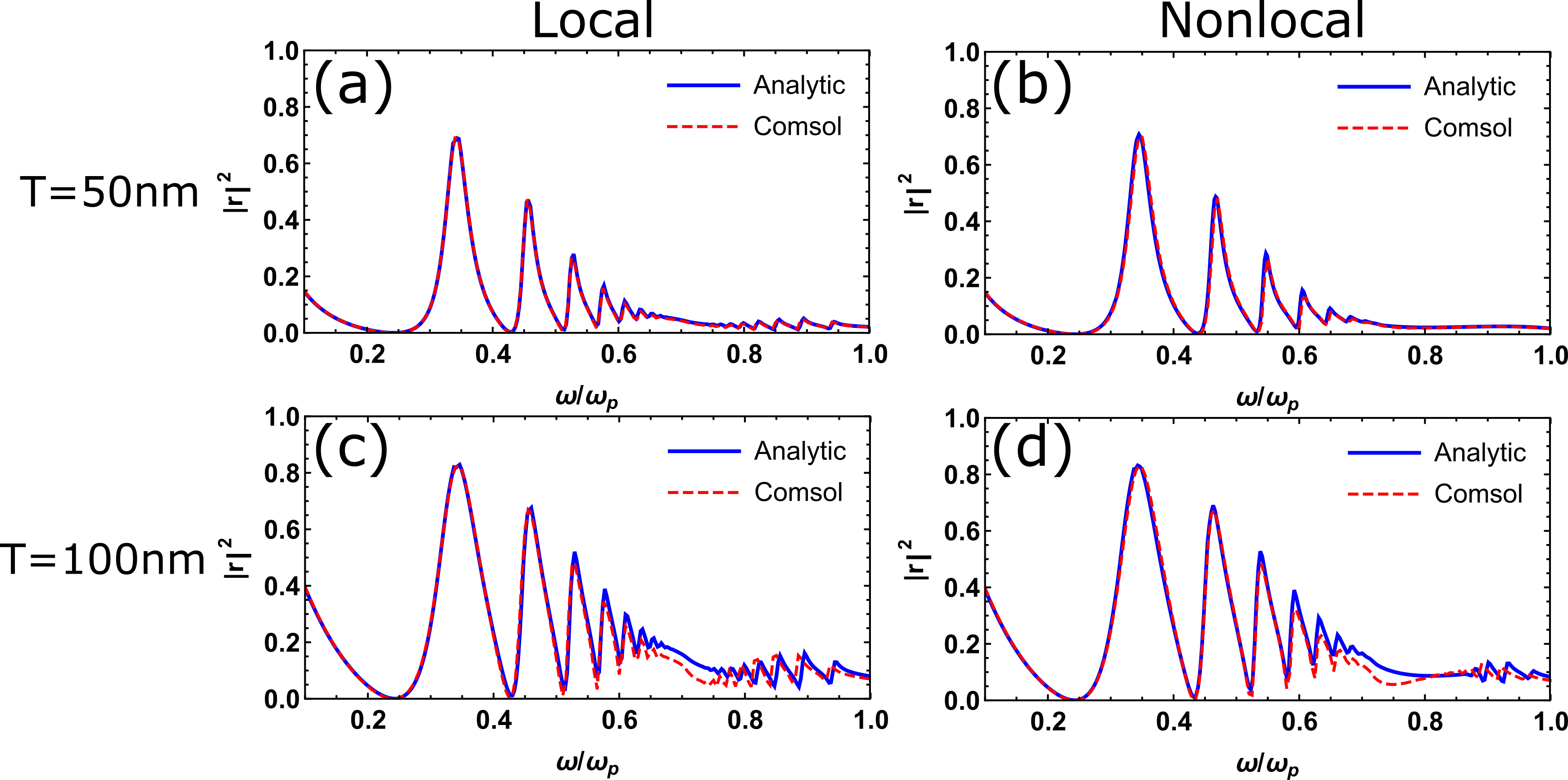}
\centering
\caption{Numerical verification of the reflection spectrum for the near-sinusoidal metasurface under normal incidence. Top row: $T=50$ nm, the maximum gap $\delta_{max} = 10$ nm and the minimum gap $\delta_{min} = 0.5$ nm; bottom row: $T=100$ nm, the maximum gap $\delta_{max} = 20$ nm and the minimum gap $\delta_{min} = 1$ nm.}
\label{NumericalVerification}
\end{figure} 

We also verify our analytic calculation with finite element solver Comsol Multiphysics. We have considered two sets of near-sinusoidal metasurfaces with period $T=50$ nm and $T=100$ nm, shown in Fig. \ref{NumericalVerification}. For a given metasurface, both local and nonlocal calculations are provided. The good agreement between analytic and numerical solutions confirms the validity of our theoretical framework. The nonlocal simulations are based on the RF and PDE modules in Comsol\citep{toscano2012modified, ciraci2012probing}, where the hydrodynamic system of equations in the metal are implemented as
\begin{equation}
\begin{split}
& \nabla \times \nabla \times \mathbf{E} = k_0^2 \mathbf{E} + i \omega \mu_0 \mathbf{J} \\
&\beta^2 \nabla (\nabla \cdot \mathbf{J}) + \omega(\omega+i \Gamma)\mathbf{J} = i\omega \omega_p^2 \varepsilon_0 \mathbf{E}
\end{split} 
\end{equation}

\bibliography{reference}

\end{document}